\documentclass{article}

\usepackage[english]{babel}
\usepackage{cite}
\usepackage[letterpaper,top=2cm,bottom=2cm,left=3cm,right=3cm,marginparwidth=1.75cm]{geometry}

\usepackage{amsmath}
\usepackage{graphicx}
\usepackage[colorlinks=true, allcolors=blue]{hyperref}
\usepackage{authblk}
\usepackage{indentfirst}
\usepackage{float}
\usepackage{booktabs}
\usepackage{multirow}
\usepackage{amsmath}
\usepackage{caption}
\usepackage{amsfonts}  
\usepackage{amssymb}   

\title{Systematic Evaluation of Stencil Configuration, Forcing Scheme, and Resolution Effects in the Stratified Taylor--Green Vortex: A Lattice Boltzmann Study}
\author[1]{Hongxuan Zhang}
\affil[1]{School of Atmospheric Sciences, Sun Yat-sen University, Zhuhai, 51900, Guangdong, China.}
\affil[1]{\textit{Email: zhanghx239@mail2.sysu.edu.cn}}
\begin{document}
\maketitle
\begin{abstract}
The rigorous simulation of stratified turbulence remains challenging due to pronounced flow anisotropy, suppressed vertical transport, and high sensitivity to numerical dissipation. This study systematically evaluates the predictive capability of the lattice Boltzmann method (LBM) for a three-dimensional stratified Taylor--Green vortex. Within a double-distribution-function framework under the Boussinesq approximation, we examine the influence of stencil configurations, forcing formulations, and spatial resolutions up to $256^3$, with validation against spectral DNS benchmarks. The results demonstrate that the D3Q27$\times$19 configuration achieves an optimal balance between numerical accuracy and computational efficiency, accurately reproducing the temporal evolution of kinetic and potential energies as well as the characteristic double-peak dissipation structure. Grid-sensitivity analysis further reveals that potential energy and fine-scale turbulent structures are significantly more resolution-dependent than kinetic energy, requiring a minimum resolution of $256^3$ for quantitative convergence. Moreover, under strongly stratified conditions, the velocity-shift forcing schemes outperform discrete source-term approaches, reducing the overall error by approximately 45.54\%. Overall, this work provides practical guidelines for high-fidelity LBM simulations of stratified turbulence and highlights that the coordinated selection of stencil isotropy, spatial resolution, and force discretization is essential for accurately capturing energy cascade and mixing dynamics.
\end{abstract}

\vspace{1em}

\section{Introduction}

Atmospheric and oceanic circulations are canonical examples of stably stratified flows, in which the density of the fluid decreases continuously as height increases\cite{lindborg2006energy,riley2000fluid}. Stable density gradients suppress vertical overturning motions\cite{riley2008stratified}, resulting in turbulent dynamics that differ fundamentally from those observed in unstratified flows\cite{diamessis2011similarity,caulfield2021layering}. A comprehensive understanding of stratified flows is of central importance to meteorology, oceanography, and planetary science. However, the vast spatial and temporal scales of these systems present an extreme challenge for physical experiments. Consequently, numerical simulation has become an essential tool for studying such complex phenomena\cite{almalkie2012kinetic}.


Previously, most numerical studies of stratified flows relied on macroscopic continuum solvers that resolve the Navier–Stokes equations under the Boussinesq approximation\cite{brethouwer2012turbulent,waite2004stratified}. Recently, the lattice Boltzmann method (LBM) has emerged as a highly effective mesoscopic alternative\cite{aidun2010lattice,he1997theory,chen1998lattice,luo2000theory}. With stream--and--collide operations at its core, the LBM is computationally efficient and provides excellent parallel performance\cite{mohamad2011lattice}. Within this framework, the double--distribution--function (DDF) formulation has gradually become a popular strategy for coupled momentum--scalar transport problems. This is because it is numerically stable and consistently recovers both the hydrodynamic and scalar transport equations\cite{shan1996simulation,peng2003simplified}. Recent comparative studies have shown that the LBM is as accurate as highly optimised finite difference solvers for weakly compressible flows\cite{khirevich2015coarse,guo2002discrete,wilde2023stratified}. Consequently, the LBM shows significant potential for simulating stable stratification.

In recent years, researchers have started to apply the LBM to stably stratified flows. For example, Kefayati et al.\cite{kefayati2019lattice} expanded and optimised the DDF model for environments involving double--diffusive convection and complex density gradients. Ottolenghi et al.\cite{ottolenghi2018lattice} simulated gravity currents by combining the LBM with a large--eddy simulation (LES) approach. This allowed them to accurately capture the complex instability features induced by the bottom wall. Furthermore, Taha et al.\cite{taha2022lattice} systematically analyzed buoyancy-driven turbulence to evaluate the method's numerical accuracy and stability in complex thermal convection. However, although the LBM has been extensively validated in isotropic turbulence and unstable thermal convection, studies of stable stratification tend to focus on simple, boundary-dominated flows or use subgrid-scale parameterisations inherent in LES. Systematic assessments of fully developed, unbounded, stably stratified turbulence using pure LBM at the direct numerical simulation (DNS) level are still lacking.

Addressing this DNS--level gap poses significant methodological challenges. The pronounced anisotropy, suppressed vertical transport, and intricate wave--turbulence interactions induced by stable stratification \cite{brethouwer2007scaling} imply that flow evolution is strongly influenced by subtle density fluctuations. This places stringent demands on numerical methods, which must accurately resolve buoyancy effects and the evolution of small--scale structures under conditions of low numerical dissipation \cite{waite2011stratified}. Key statistical quantities in stratified flows are highly sensitive to discretization errors; if the numerical configuration is inappropriate --for instance, due to inaccuracies in scalar transport discretization or buoyancy forcing treatment--nonphysical artifacts may accumulate over long-term integration, thereby obscuring the underlying physical dynamics \cite{guo2002discrete,lallemand2000theory,chai2013lattice}. This underscores the need for dedicated benchmark studies to systematically evaluate the predictive capability and numerical limitations of the LBM in such flows \cite{garg2000stably}.

The three-dimensional Taylor--Green vortex (TGV) has been widely adopted as a canonical benchmark for evaluating the dissipation characteristics and stability of numerical algorithms \cite{brachet1983small,gassner2013accuracy,kajzer2014large}. Owing to its analytically prescribed initial conditions and reproducible flow evolution, the TGV is frequently employed to compare the performance of various numerical methods \cite{van2011comparison}. In recent years, this configuration has been extended to supersonic and stratified flow regimes, providing a standardized testing platform for investigating complex flow dynamics. Under unstratified conditions, the initial large-scale modes of the TGV eventually evolve into nearly homogeneous turbulence \cite{bull2014simulation}. In contrast, under stable stratification, buoyancy effects substantially modify the flow evolution, confining mixing predominantly to horizontal planes \cite{rosenberg2015evidence}. These contrasting dynamical features render the stratified Taylor--Green vortex (STGV) an ideal benchmark for assessing the capability of numerical methods to predict stratified turbulence.

Based on the above considerations, the present study adopts the three-dimensional STGV as a rigorous benchmark. The primary objective is to identify an optimal numerical configuration for simulating stratified turbulence within the DDF--LBM framework. To this end, a series of systematic cross--comparative simulations are conducted to evaluate the performance of different combinations of discrete velocity stencils, grid resolutions, and forcing schemes under weak buoyancy forcing. Key statistical quantities are monitored to quantitatively assess the influence of these numerical choices on solution accuracy.

The remainder of this work is organized as follows: Section 2 outlines the numerical methodology, covering the governing equations, the LBM framework, and the initial conditions and relevant parameters of the TGV. Section 3 provides a detailed analysis and discussion of the results. Finally, Section 4 summarizes the key conclusions.

\section{Methodology}

The fluid flow is governed by the incompressible Navier--Stokes equations under the Boussinesq approximation\cite{vallis2017atmospheric,brethouwer2007scaling,turner1979buoyancy}. The conservation of mass and momentum can be expressed in tensor notation as follows:
\begin{equation}
    \partial_\alpha u_\alpha= 0,
    \label{eq:continuity}
\end{equation}
\begin{equation}
    \partial_t u_\alpha + u_\beta \partial_\beta u_\alpha = -\frac{1}{\rho_0}\partial_\alpha p - \frac{\rho'}{\rho_0} G \delta_{\alpha z} + \nu \partial_\beta \partial_\beta u_\alpha,
\end{equation}
where $u_\alpha$ represents the velocity component, $p$ is the pressure, $G$ is the gravitational acceleration and $\nu$ denotes the kinematic viscosity.
The density perturbation $\rho^{\prime}$ is defined as the part of the total density that fluctuates. It is described by the following equation:
\begin{equation}
\rho(x,y,z,t) = \rho_0 + \rho_b(z) + \rho'(x,y,z,t).
\label{eq:density-decomposition}
\end{equation}
Here $\rho_0$ is a constant reference density and $\rho_b(z)$ represents the background stratification, which decreases linearly as the vertical coordinate $z$ increases. The density perturbation's temporal evolution is governed by
\begin{equation}
    \partial_t \rho' + u_\alpha \partial_\alpha \rho' = -u_z \frac{\mathrm{d} \rho_b}{\mathrm{d} z} + \tilde{D} \partial_\alpha \partial_\alpha \rho',
\end{equation}
where $\tilde{D}$ stands for the mass diffusivity.

\subsection{Approximation of the momentum equation}
The governing fluid dynamics are resolved using the DDF framework. The evolution of the discrete particle distribution function, $f_i(\mathbf{x}, t)$, is governed by the standard lattice Boltzmann equation \cite{qian1992lattice,chen1998lattice}:
\begin{equation}
    f_i(\mathbf{x} + \mathbf{e}_i\delta_t, t + \delta_t) = f_i(\mathbf{x}, t) + \Omega_i^f + \mathbf{F}_i,
\end{equation}
where $\mathbf{e}_i$ is the discrete velocity in the $i$-th direction, $\delta_t$ is the time step, and $\mathbf{F}_i$  represents the discrete source term incorporating external forces. The macroscopic fluid density $\rho$ and the velocity $\mathbf{u}$ are derived from the moments of the distribution function, taking into account the discrete lattice effects:
\begin{equation}
    \rho(\mathbf{x},t) = \sum_i f_i(\mathbf{x},t),
\end{equation}
\begin{equation}
    \mathbf{u}(\mathbf{x},t) = \frac{1}{\rho} \sum_i \mathbf{e}_i f_i(\mathbf{x},t) + \frac{\mathbf{F} \delta_t}{2\rho},
\end{equation}

For the collision term $\Omega_i^f$, the Bhatnagar--Gross--Krook approximation is adopted. This operator relaxes the distribution function towards a local equilibrium state $f_i^{\mathrm{eq}}$ at a rate determined by the relaxation time $\tau_f$:
\begin{equation}
    \Omega_i^f = -\frac{1}{\tau_f} (f_i - f_i^{\mathrm{eq}}).
\end{equation}
Assuming standard lattice units, the relaxation parameter is connected to the kinematic viscosity via the relationship $\tau_f = \nu/c_s^2 + 0.5$, where $c_s$ is the lattice speed of sound. The equilibrium distribution function is obtained via a second-order Hermite expansion of the Maxwell--Boltzmann distribution \cite{he1997theory}:
\begin{equation}
    f_i^{\mathrm{eq}}(\rho, \mathbf{u}) = \rho w_i \left( 1 + \frac{\mathbf{e}_i \cdot \mathbf{u}}{c_s^2} + \frac{(\mathbf{e}_i \cdot \mathbf{u})^2}{2c_s^4} - \frac{\mathbf{u}^2}{2c_s^2} \right).
\end{equation}

In this work, the classical Shan--Chen forcing scheme is primarily utilized to handle the buoyancy forces. Unlike schemes that introduce an explicit forcing term on the right-hand side of Eq. (5), the Shan--Chen approach incorporates body forces implicitly by shifting the macroscopic velocity entering the equilibrium distribution.  Consequently, the modified equilibrium velocity $\mathbf{u}^{\mathrm{eq}}$ inputted into Eq.~(9) is defined as follows \cite{shan1993lattice}:
\begin{equation}
    \mathbf{u}^{\mathrm{eq}}(\mathbf{x},t) = \frac{1}{\rho} \sum_i \mathbf{e}_i f_i(\mathbf{x},t) + \frac{\mathbf{F} \tau_f}{2\rho}.
\end{equation}

\subsection{Approximation of the advection-diffusion equation}

The evolution of the macroscopic density perturbation $\rho'$ is governed by the advection-diffusion equation. In the DDF framework, this is resolved by introducing a second set of discrete distribution functions, $g_i(\mathbf{x}, t)$, whose evolution is described by a lattice Boltzmann equation that includes a discrete source term $S_i$ to account for the effects of stratification\cite{shan1996simulation}::
\begin{equation}
    g_i(\mathbf{x} + \mathbf{e}_i\delta_t, t + \delta_t) = g_i(\mathbf{x}, t) + \Omega_i^g + S_i,
\end{equation}
where the source term $S_i$ depends on the local vertical velocity $u_z$ and the background density gradient:
\begin{equation}
    S_i = w_i \left( -u_z \frac{\mathrm{d}\rho_b}{\mathrm{d}z} \right)_{\mathrm{LU}}.
\end{equation}
The subscript "LU" indicates that the variables must be converted to lattice units based on the grid spacing $\delta_x$, defined by the following conversion relation:
\begin{equation}
    \left( \frac{\mathrm{d}\rho_b}{\mathrm{d}z} \right)_{\mathrm{LU}} = \left( \frac{\mathrm{d}\rho_b}{\mathrm{d}z} \right) \delta_x.
\end{equation}

Similar to the fluid solver, the collision operator for the scalar field employs the BGK approximation with a scalar relaxation time $\tau_g$:
\begin{equation}
    \Omega_i^g = -\frac{1}{\tau_g} (g_i - g_i^{\mathrm{eq}}).
\end{equation}
The relaxation parameter $\tau_g$ is determined by the mass diffusivity $\tilde{D}$ through the relation $\tau_g = \tilde{D}/c_s^2 + 0.5$. The corresponding equilibrium distribution function $g_i^{\mathrm{eq}}$ is defined as:
\begin{equation}
    g_i^{\mathrm{eq}}(\rho', \mathbf{u}) = (\rho' + \rho_0) \Theta_i^{\mathrm{eq}}(\mathbf{u}),
\end{equation}
where $\Theta_i^{\mathrm{eq}}(\mathbf{u}) = f_i^{\mathrm{eq}} / \rho$ represents the weight polynomial dependent on the local velocity. While the discrete velocity set for $g_i$ can differ from that of $f_i$, adopting identical stencils simplifies the implementation and optimizes computational efficiency by allowing the direct reuse of the term $\Theta_i^{\mathrm{eq}}$.

The macroscopic density perturbation $\rho'$ is determined from the zeroth moment of $g_i$. To minimize discrete errors associated with the source term treatment, a correction term is incorporated into the macroscopic calculation \cite{chai2013lattice}:
\begin{equation}
    \rho'(\mathbf{x}, t) = \sum_i g_i(\mathbf{x}, t) + \frac{1}{2} \sum_i R_i - \rho_0,
\end{equation}
\begin{equation}
    R_i = \left( 1 - \frac{1}{2\tau_g} \right) S_i.
\end{equation}

The coupling between the scalar field and the fluid flow is ultimately achieved through the Boussinesq approximation. The density perturbation $\rho'$ acts as an active scalar, generating a local buoyancy force $\mathbf{F}$ that feeds back into the fluid's momentum equation:
\begin{equation}
    \mathbf{F} = 
    \begin{pmatrix}
    0 \\
    0 \\
    -G\rho'
    \end{pmatrix}.
\end{equation}

It is worth noting that this DDF framework explicitly decouples the fluid density $\rho$ (which is related to the hydrodynamic pressure via the equation of state) from the density perturbation $\rho'$. This separation is physically justified, as the effects of compressive work and viscous heating are completely negligible within the incompressible, Boussinesq flow regime under consideration.

\subsection{Stencils Model}

The choice of discrete velocity stencil is a central factor governing the accuracy and efficiency of LBM simulations \cite{rubinstein2008theory,philippi2006continuous}. Beyond a implementation detail, the lattice stencil determines the streaming pathways of the particle distribution functions, the recoverable order of macroscopic moments, and the degree of numerical isotropy. It reveals the balance between achieving a highly accurate solution and the associated computational cost \cite{d2002multiple,shan2006kinetic}.

Since the computational cost of LBM increases approximately linearly with the number of discrete velocities, practical implementations often require a compromise. Higher--order stencils with improved isotropy are typically employed for the momentum equations to ensure hydrodynamic accuracy \cite{peng2003simplified}, whereas more compact stencils are frequently adopted for scalar transport to reduce computational overhead. However, in stably stratified flows, the dynamics can be strongly influenced by subtle density variations and weak buoyancy imbalances, increasing sensitivity to numerical dissipation and discretization errors. Although reduced--order scalar stencils significantly lower computational cost \cite{peltier2003mixing}, their limited set of discrete directions may compromise the accurate representation of buoyancy--driven effects and enhance spurious numerical artifacts \cite{yoshida2010multiple}.


To address this gap and identify a suitable balance between efficiency and accuracy, this work conducts a systematic comparison of multiple stencil combinations. Four commonly used velocity sets---D3Q7, D3Q15, D3Q19, and D3Q27---are employed to construct various coupled fluid--scalar configurations. Their predictive performance is evaluated within the STGV benchmark. Figure \ref{fig:stencils} and Table A.1 show the geometric structures and detailed parameters of the adopted lattice stencils.
\begin{figure}[H]
    \centering
    \includegraphics[width=0.8\linewidth]{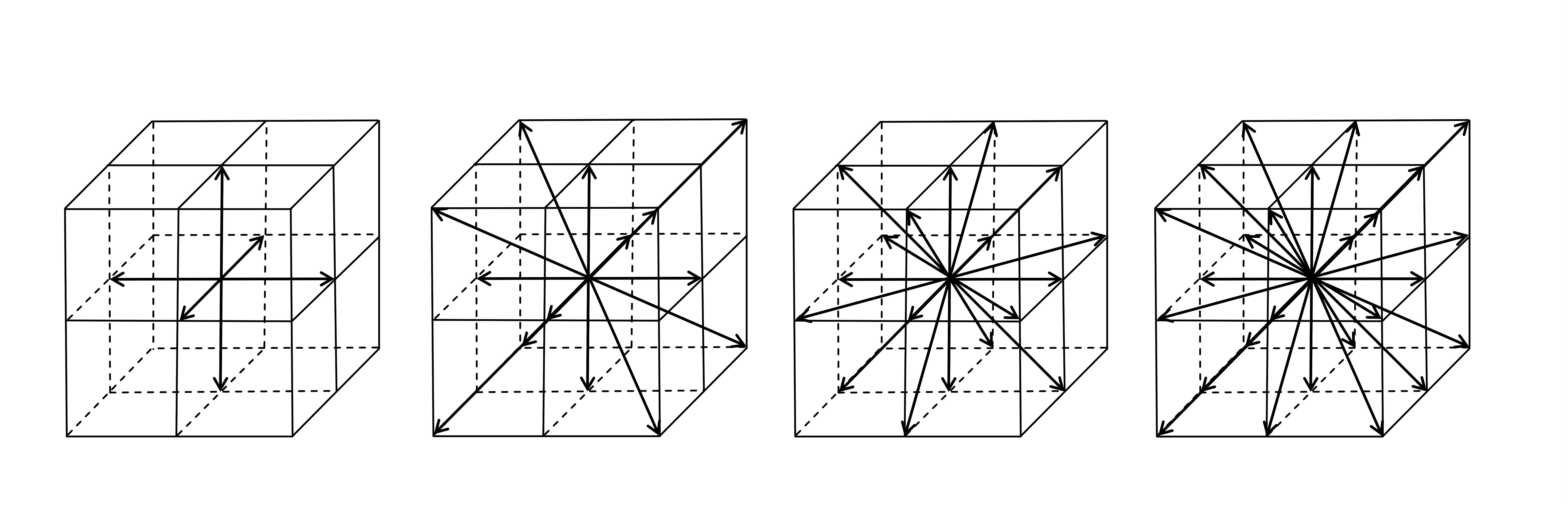}
    \caption{Lattice velocity stencils used in this study. From left to right: D3Q7, D3Q15, D3Q19 and D3Q27.}
    \label{fig:porosity_effect_flux}
\end{figure}

The notation D3Q$m \times n$ is used to describe the multiphysics coupling strategy, where $m$ denotes the number of discrete velocities for the hydrodynamic distribution function $f_i$, and $n$ corresponds to that for the scalar distribution function $g_i$. For instance, D3Q27$\times$19 represents a 27-velocity fluid solver coupled with a 19-velocity scalar model, whereas D3Q19$\times$15 denotes a 19-velocity fluid solver combined with a 15-velocity scalar solver.

\begin{table}[H]
    \caption{Compilation of the stencils D3Q7, D3Q15, D3Q19, and D3Q27.}
    \label{tab:stencils_all}
    \renewcommand{\arraystretch}{1.3} 
    \begin{tabular}{lllll}
        \toprule
        Stencil & Lattice speed of sound $c_s$ & Index $i$ & Weight  $w_i$ & Discrete velocity $\mathbf{e}_i$\\
        \midrule
        
        \multirow{2}{*}{D3Q7} & \multirow{2}{*}{$1/2$} 
          & $0$ & $1/4$ & $(0,0,0)$ \\
          & & $1,\dots,6$ & $1/8$ & $(\pm 1, 0, 0), (0, \pm 1, 0), (0, 0, \pm 1)$ \\
        \midrule
        
        \multirow{3}{*}{D3Q15} & \multirow{3}{*}{$1/\sqrt{3}$} 
          & $0$ & $2/9$ & $(0,0,0)$ \\
          & & $1,\dots,6$ & $1/9$ & $(\pm 1, 0, 0), (0, \pm 1, 0), (0, 0, \pm 1)$ \\
          & & $7,\dots,14$ & $1/72$ & $(\pm 1, \pm 1, \pm 1)$ \\
        \midrule
        
        \multirow{3}{*}{D3Q19} & \multirow{3}{*}{$1/\sqrt{3}$} 
          & $0$ & $1/3$ & $(0,0,0)$ \\
          & & $1,\dots,6$ & $1/18$ & $(\pm 1, 0, 0), (0, \pm 1, 0), (0, 0, \pm 1)$ \\
          & & $7,\dots,18$ & $1/36$ & $(\pm 1, \pm 1, 0), (\pm 1, 0, \pm 1), (0, \pm 1, \pm 1)$ \\
        \midrule
        
        \multirow{4}{*}{D3Q27} & \multirow{4}{*}{$1/\sqrt{3}$} 
          & $0$ & $8/27$ & $(0,0,0)$ \\
          & & $1,\dots,6$ & $2/27$ & $(\pm 1, 0, 0), (0, \pm 1, 0), (0, 0, \pm 1)$ \\
          & & $7,\dots,18$ & $1/54$ & $(\pm 1, \pm 1, 0), (\pm 1, 0, \pm 1), (0, \pm 1, \pm 1)$ \\
          & & $19,\dots,26$ & $1/216$ & $(\pm 1, \pm 1, \pm 1)$ \\
        \bottomrule
    \end{tabular}
\end{table}

\subsection{Stratified Taylor--Green vortex}
This work adopts a three--dimensional STGV as the benchmark case. This case is based on the classical Taylor--Green vortex described by Brachet et al.\cite{brachet1983small,jadhav2021assessment,bedrunka2021lettuce}, but with the addition of density stratification.

The dynamics of the STGV are governed by three dimensionless numbers: the Reynolds number $Re_0 = UL/\nu$, the Froude number $Fr_0 = U/(NL)$, and the Schmidt number $Sc = \nu/\tilde{D}$. In the present simulations, the characteristic velocity and length scales are set to $U = L = 1$. Across all test cases, $Re_0$ is fixed at $1600$ and $Sc$ is maintained at $0.7$. Unless specifically stated otherwise to examine particular parametric effects, the Froude number is uniformly set to $Fr_0 = 1$. The buoyancy frequency $N$ is determined by the background density gradient and gravity:

\begin{equation}
N=\sqrt{-\frac{G}{\rho_{0}} \frac{\partial \rho_{b}}{\partial z}} .
\end{equation}

The computational domain is a three--dimensional periodic cubic box $V=[0,2\pi]^3$. The initial velocity and pressure fields are based on an analytical TGV structure:
\begin{equation}
\mathbf{u}(\mathbf{x}, t = 0) = U
\begin{pmatrix}
\sin(x)\cos(y)\cos(z) \\
-\cos(x)\sin(y)\cos(z) \\
0
\end{pmatrix},
\label{eq:u0}
\end{equation}

\begin{equation}
p(x,t=0) = \frac{1}{16}\,\bigl(\cos(2x) + \cos(2y)\bigr)\cos(2z+2),
\label{eq:initial-pressure}
\end{equation}
with the initial density perturbation set to $\rho'(\mathbf{x}, t = 0) = 0.$ Time is non-dimensionalized using the convective timescale $t_c=L/U=1.$
In the weakly compressible LBM framework, the Mach number $Ma = U_{LB}/c_s$ relate the lattice characteristic velocity to the lattice sound speed. This study adopts a mach number of 0.1, corresponding to a lattice velocity $U_{LB} = 0.0577$, which is used to initialize the velocity field.

\section{Results and discussion}

\subsection{Model Validation}
The present LBM solver is first validated against the classical unstratified TGV benchmark. For this validation, the fluid density is assumed constant, and the buoyancy term is ignored. The flow field is initialized using the analytical TGV solution introduced earlier \cite{taylor1937mechanism}.

All simulations in this section adopt the D3Q27 lattice model. The three--dimensional computational domain is uniformly discretized with a grid resolution of \(256^3\) grid points. To satisfy the incompressible limit under the weakly compressible LBM framework, the Mach number is set to \(Ma=0.05\). In addition, to comprehensively assess the performance of the numerical solver, simulations are carried out over a broad range of Reynolds numbers (\(Re \in \{100,200,400,800,1600,3000\}\)), covering laminar, transitional, and fully turbulent flow regimes.

\begin{figure}[H]
    \centering
    \includegraphics[width=0.6\linewidth]{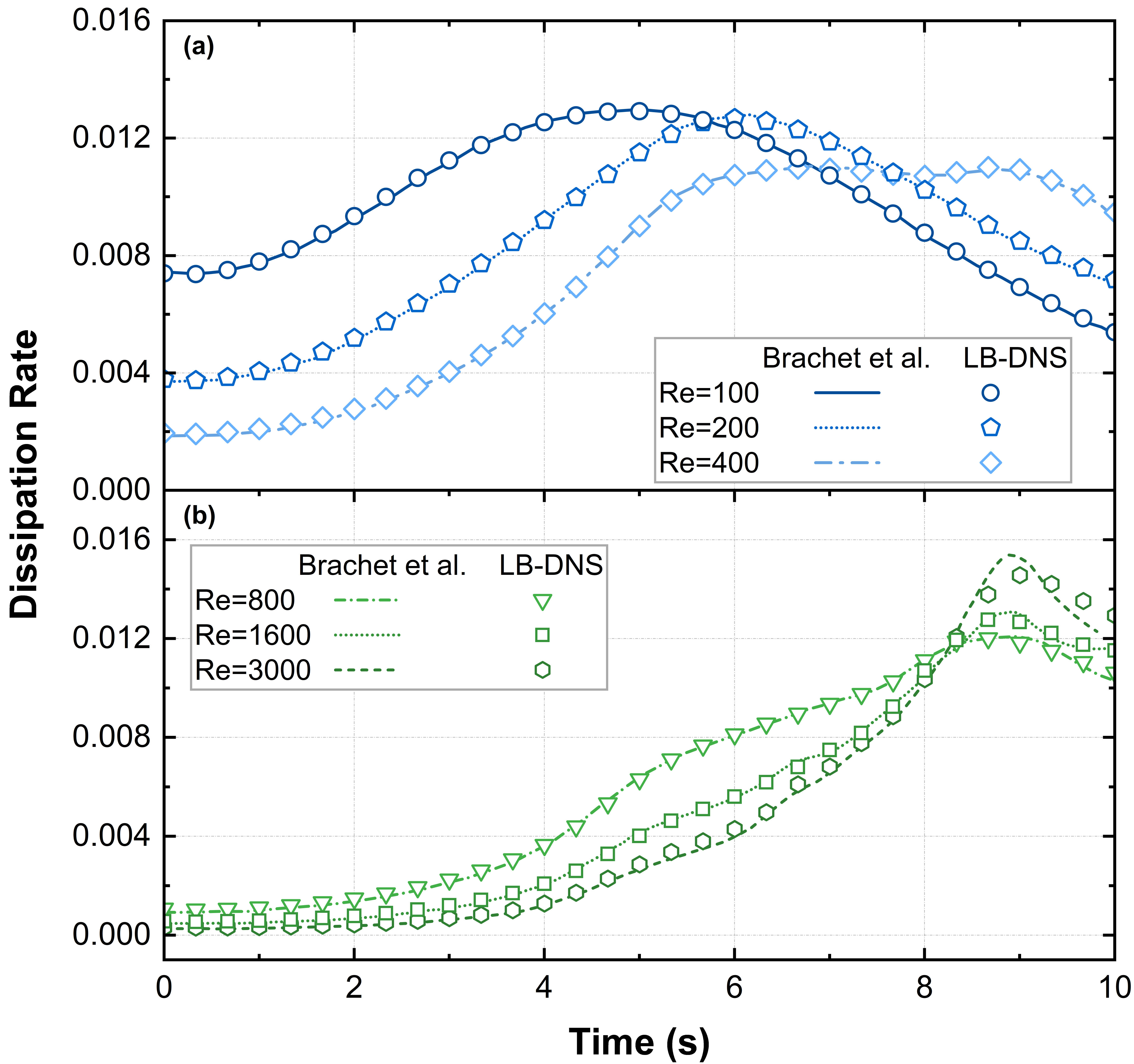}
    \caption{
     Temporal evolution of the kinetic energy dissipation rate $\varepsilon(t)$ for the unstratified TGV at various Reynolds numbers. The present LBM results show good agreement with the DNS data of Brachet et al. \cite{brachet1983small}.
     }
    \label{fig:taylor_green_energy_dissipation}
\end{figure}

Validation is conducted by examining the temporal evolution of the kinetic energy dissipation rate. First, the volume--averaged turbulent kinetic energy $E_k$ is calculated from the resolved velocity field. The energy dissipation rate is subsequently derived using the finite--difference approximation \(\varepsilon(t) \approx -\mathrm{d}E_\mathrm{k}/\mathrm{d}t\), where a second--order central difference scheme is adopted for time--derivative evaluation. Fig.~\ref{fig:taylor_green_energy_dissipation} compares the present LBM predictions with high--fidelity spectral direct numerical simulation (DNS) data reported by Brachet et al.\cite{brachet1983small}.

The result shows that the D3Q27-based LBM solver is highly accurate for $Re \leq 1600$. The numerical scheme precisely captures both the timing and the magnitude of the peak dissipation rate, as well as the subsequent viscous decay process. At the highest Reynolds number of $Re = 3000$, a slight underprediction of the peak dissipation rate is observed. This deviation is attributed to the limited spatial resolution of the $256^3$ grid, which under--resolves the smallest dissipative scales at such a high $Re$\cite{gassner2013accuracy}. Nevertheless, the overall temporal evolution closely aligns with the spectral DNS benchmark.

In summary, the D3Q27 lattice model has been verified as both numerically stable and accurate for simulating the unstratified TGV. This validation confirms the reliability of the present solver and establishes a solid foundation for subsequent numerical studies of stratified flows and coupled scalar fields. With this validated hydrodynamic baseline established, the following sections introduce density stratification to investigate the evolution of potential energy $E_p$ and internal wave dynamics, where the choice of velocity stencils and adequate spatial resolution will become even more critical.

\subsection{ Kinetic, potential and total energy}


The impact of different stencil combinations on energy transfer in the STGV can be assessed by tracking three key macroscopic quantities over time: volume--averaged kinetic energy $E_k$, potential energy $E_p$, and total energy $E_{total}$. These three quantities are defined as follows:

\begin{equation}
E_k = \frac{1}{(2\pi)^3} \int_V \frac{1}{2} |\boldsymbol{u}|^2 \, dV,
\end{equation}

\begin{equation}
E_{\mathrm{p}} = \frac{1}{(2\pi)^3} \int_V \frac{1}{2} \left( \frac{G}{N} \frac{\rho'}{\rho_0} \right)^2 dV,
\end{equation}

\begin{equation}
E_{total}=E_k+E_p.
\end{equation}

To investigate the influence of coupling different discrete velocity stencil on the energy evolution, four representative configurations are considered on a $128^3$ computational grid: D3Q27$\times$19, D3Q19$\times$15, D3Q15$\times$7, and D3Q7$\times$27. The first three form a sequence of progressively lower-order discretizations applied simultaneously to both the momentum and scalar fields. This hierarchy allows the sensitivity of energy evolution to stencil order to be quantified. The D3Q7$\times$27 case is additionally included to assess whether increasing the scalar stencil order alone can compensate for errors arising from lower-order momentum discretizations. All results are compared against spectral DNS data from Jadhav et al.\cite{jadhav2021assessment}.

\begin{figure}[H]
    \centering
    \includegraphics[width=0.7\linewidth]{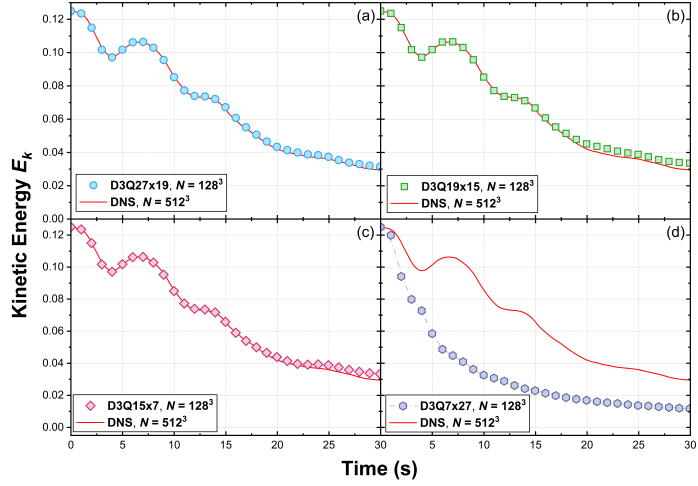}
    \caption{The evolution of the kinetic energy $E_k$ over time for four LBM configurations. DNS from Jadhav et al.\cite{jadhav2021assessment}}
    \label{fig:Ek_stencils}
\end{figure}


First, the evolution of the kinetic energy $E_k$ is examined, as shown in Fig.~\ref{fig:Ek_stencils}. The D3Q27$\times$19 configuration accurately captures both the initial decay and growth of kinetic energy and the appearance and subsequent decay of its peak. It also remains highly consistent with the reference throughout the simulation. And the D3Q19\(\times\)15 and D3Q15\(\times\)7 medium-order configurations both fit the DNS curve acceptably well in the \(t \lesssim 20\) stage. There are only minor deviations from the reference in terms of peak magnitude and occurrence time. However, as the simulation progresses, both configurations exhibit a distinct tendency toward under--dissipation. Their kinetic energy $E_k$ decay rates are notably slower than those predicted by DNS. This results in an ever--increasing difference at later stages. Conversely, the D3Q7$\times$27 configuration exhibits remarkably poor performance. From the outset, this configuration shows excessive numerical dissipation, triggering a rapid and premature energy decay while severely underpredicting the peak value. These results clearly demonstrate that the low-order D3Q7 stencil is fundamentally inadequate for resolving the momentum equations in this context.

\begin{figure}[H]
    \centering
    \includegraphics[width=0.7\linewidth]{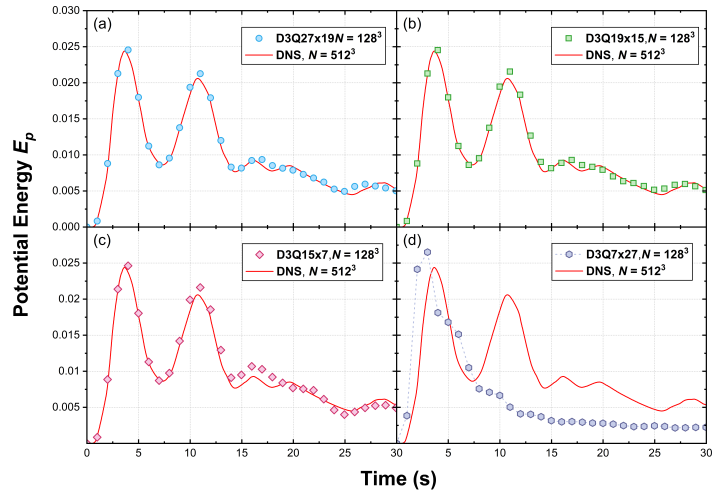}
    \caption{The evolution of the potential energy $E_p$ over time for four LBM configurations. DNS from Jadhav et al.\cite{jadhav2021assessment}}
    \label{fig:Ep_stencils_128}
\end{figure}

Fig.~\ref{fig:Ep_stencils_128} compares the evolution of the potential energy $E_p$ for different configurations.  In the early stages, the $E_p$ profiles obtained from all configurations are generally consistent with the DNS data. However, significant discrepancies emerge when $t\geq11.75$. Even for the D3Q27$\times$19 configuration, which exhibits the best overall performance, the peak magnitude and subsequent decay rate of $E_p$ differ significantly from the DNS reference. This indicates that the potential energy $E_p$ is much more sensitive to spatial resolution than the kinetic energy $E_k$. At the relatively coarse resolution of $128^3$, all configurations struggle to adequately resolve the contributions of internal gravity waves and small-scale structures to the overall potential energy dynamics. Consequently, the prediction errors at this resolution are not primarily governed by the choice of stencil configuration, but are instead dominated by the insufficient grid resolution of these fine--scale flow features.

\begin{figure}[H]
    \centering
    \includegraphics[width=0.7\linewidth]{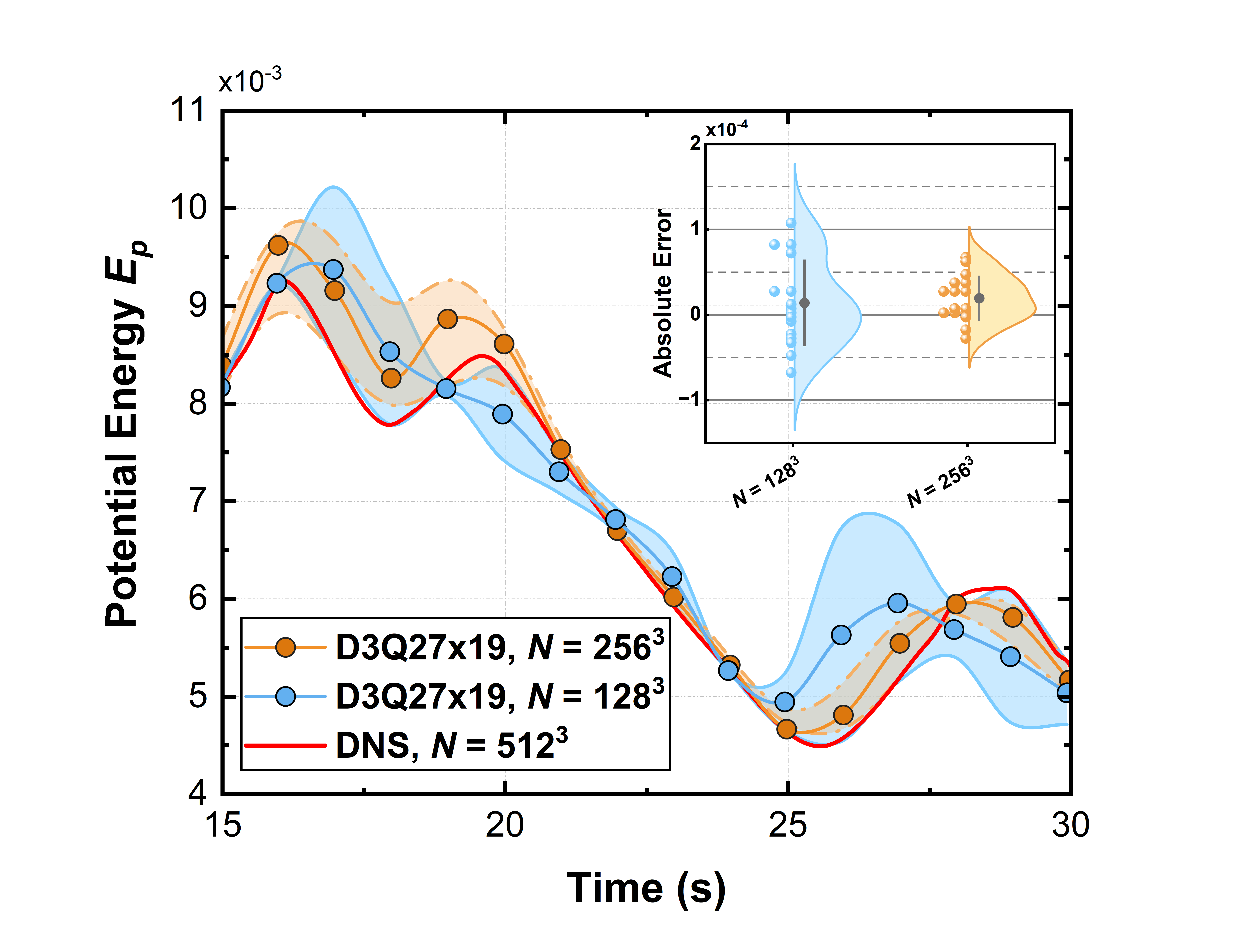}
    \caption{A comparison of the potential energy $E_p$ for different resolutions $N_g$, using the D3Q27$\times$19. DNS from Jadhav et al.\cite{jadhav2021assessment}}
    \label{fig:Ep_resolution}
\end{figure}

To further investigate the resolution dependence, a grid refinement analysis is carried out for potential energy $E_p$, as shown in Fig.~\ref{fig:Ep_resolution}. As the grid resolution increases from \(128^3\) to \(256^3\), the potential energy $E_p$ profile predicted by the D3Q27$\times$19 configuration exhibits markedly improved agreement with DNS data. Discrepancies in both peak amplitude and temporal position are substantially reduced. Quantitative error analysis indicates that the relative error of the \(E_\mathrm{p}\) profile at \(256^3\) resolution decreases by approximately 30.07\% relative to the \(128^3\) case. Evidently, optimizing only the configuration is insufficient to accurately capture the temporal evolution of potential energy $E_p$ in stratified flows; this must be combined with sufficient spatial resolution to reliably resolve vertical stratification and small--scale turbulent structures.

\begin{figure}[H]
    \centering
    \includegraphics[width=0.7\linewidth]{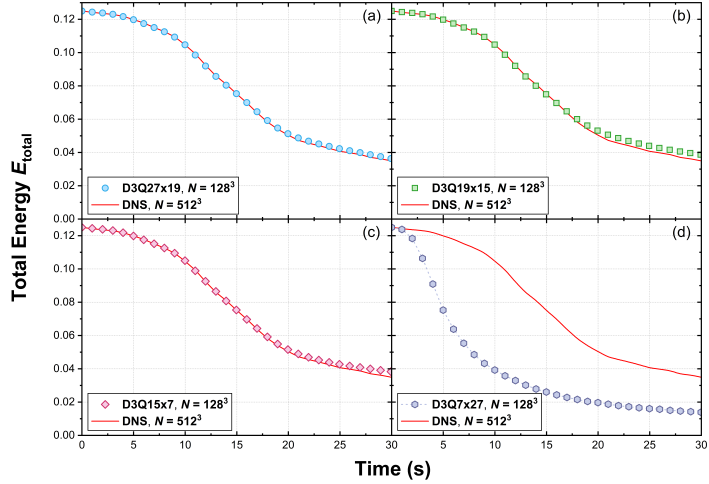}
    \caption{The evolution of the total energy $E_{total}$ over time for four LBM configurations. DNS from Jadhav et al.\cite{jadhav2021assessment}}
    \label{fig:Etot_stencils}
\end{figure}

Fig.~\ref{fig:Etot_stencils} presents how the total energy $E_{total}$ evolves over time. The observed trends are fully consistent with the preceding analyses of kinetic and potential energies. Among all configurations, D3Q27\(\times\)19 configuration maintains excellent agreement with the DNS reference throughout the entire simulation. The medium--order configurations D3Q19\(\times\)15 and D3Q15\(\times\)7 show favorable agreement with DNS data at early times, yet gradually diverge from reference results in the long--term decay stage. By contrast, the D3Q7\(\times\)27 configuration suffers from prominent spurious numerical dissipation from the onset, leading to severe underprediction of total energy throughout the entire simulation duration.

Overall, these findings indicate that an accurate discretisation of the momentum equations in the STGV requires a sufficiently high--order velocity stencil. Among the four configurations considered, D3Q27$\times$19 achieves the best agreement with the DNS reference and represents a well--balanced choice in terms of numerical accuracy and computational efficiency. The D3Q19$\times$15 and D3Q15$\times$7 configurations initially show good agreement with DNS reference, but diverge significantly during the long--term decay stage. By contrast, excessive numerical dissipation is evident from the outset of the D3Q7$\times$27 combinations. This suggests that the D3Q7 momentum stencil is not suitable for this type of stratified turbulent flow.

Finally, the resolution analysis of the potential energy shows that selecting an appropriate velocity configuration alone is insufficient for flows involving strong density stratification and internal--wave dynamics. Adequate grid resolution is equally indispensable to ensure the numerical fidelity of both the kinetic and potential energy fields.

\subsection{Dissipation of kinetic and potential energy}
This section investigates the influence of high--order stencil configurations and grid resolution on the energy dissipation characteristics of the STGV. The objective is to identify a configuration that provides an optimal balance between numerical accuracy and computational efficiency, thereby offering guidance for subsequent STGV simulations.

The total energy dissipation rate $-dE_{total}/dt$ is chosen as the main indicator. The time derivative of $E_{total}$ is computed using the second--order central difference scheme. This indicator is highly sensitive to discretisation errors in momentum and scalar transport. It can be used to characterise the total amount of kinetic and potential energy dissipated to small scales via viscosity and molecular diffusion. Based on this, the simulation results for three configurations of high-order discrete velocity templates are compared: D3Q27$\times$19, D3Q27$\times$27 and D3Q19$\times$19. 

\begin{figure}[H]
    \centering
    \includegraphics[width=0.7\linewidth]{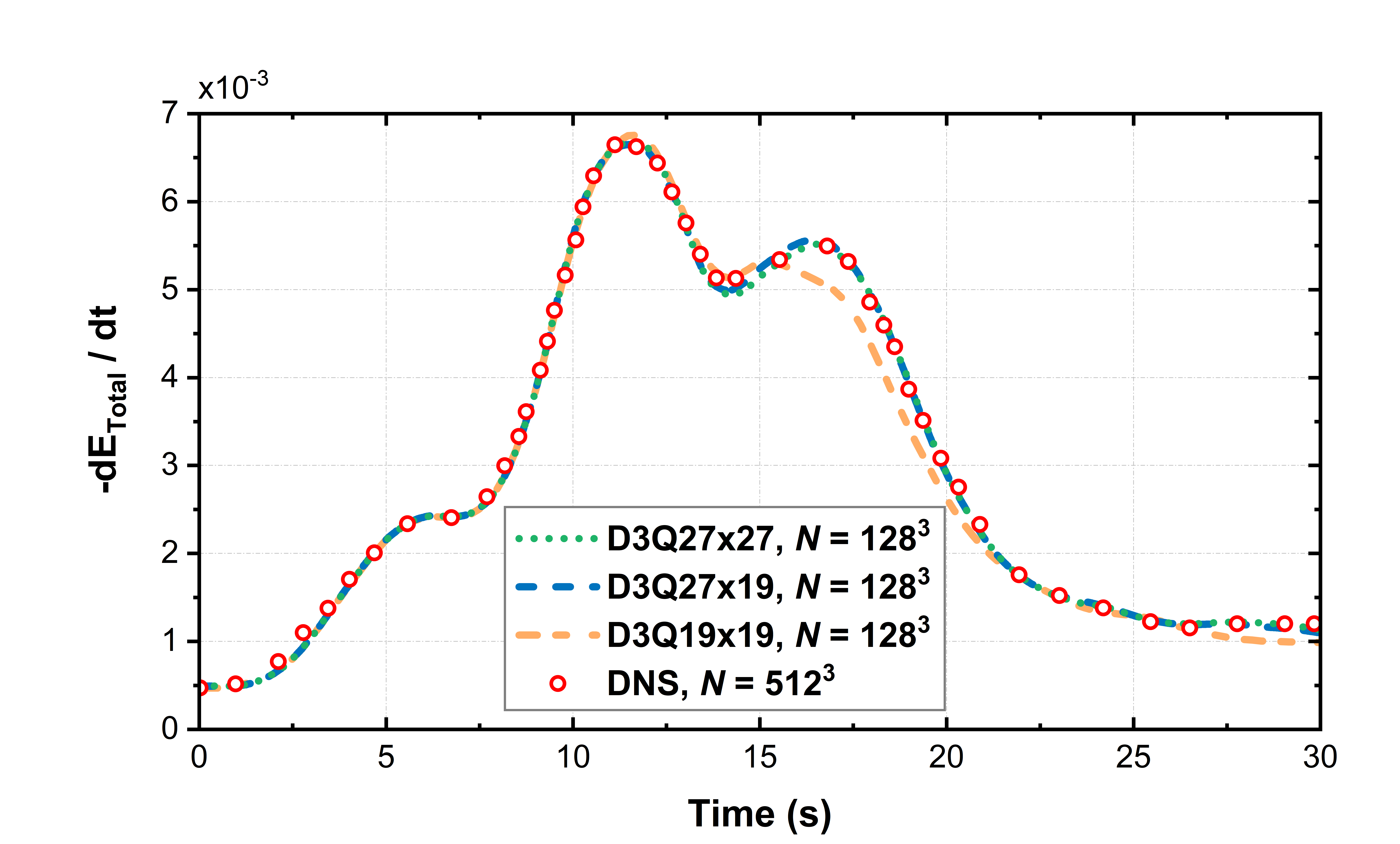}
    \caption{Time evolution of the total energy dissipation rate, $-dE_{total}/dt$, for different stencil combinations in the STGV, compared with the spectral DNS benchmark at $N_g=512^3$.}
    \label{fig:dkdt}
\end{figure}

As shown in Fig.~\ref{fig:dkdt},all three configurations accurately reproduce the dominant timescales of the energy cascade and dissipation processes in STGV. The simulation results show two distinct dissipation peaks near times $t\approx11.47$ and $t\approx16.46$. The breakdown of the large--scale Taylor--Green mode is reflected in these. This is followed by a rapid cascade of energy towards smaller scales. After these peaks, the dissipation rate gradually decreases as the flow enters its decay stage.

The three configurations produce significantly different results. The D3Q27$\times$19 configuration has the best proformence. The magnitudes and occurrence times of the two dissipation peaks are consistent with the DNS results. Furthermore, the curves during the rising and decaying phases have a similar shape. The strength of dissipation, its temporal evolution and its variation rate can all be accurately predicted. The D3Q27$\times$27 configuration, the highest--order model, shows a consistent overall trend with the D3Q27$\times$19 combination. However, minor discrepancies were noticed at the second dissipation peak. By contrast, the D3Q19$\times$19 combination shows the most huge differences. The height of the second dissipation peak is underestimated by a large margin. This has resulted in a marked error with the  DNS result. Using the total energy dissipation rate as the evaluation metric, the D3Q27$\times$19 configuration achieves a level of precision comparable to the D3Q27$\times$27 configuration and superior to the D3Q19$\times$19 configuration.

In terms of computational cost, the D3Q27$\times$27 configuration employs a 27--velocity stencil for both the momentum and the scalar field. This means that more distribution functions need updating. Consequently, the D3Q27$\times$27 configuration requires more memory and involves greater communication overhead than the D3Q27$\times$19 configuration. However, the improvement in numerical accuracy is hardly noticeable. Although the D3Q19$\times$19 configuration requires less computation, it is less effective at capturing both the dissipation peaks and the late decay stage. Overall, its performance is inferior to that of the D3Q27$\times$9 configuration.

For the 'fully high--order' stencil, the D3Q27$\times$19 configuration offers the greatest accuracy while ensuring computational efficiency, given the current parameters and mesh settings. This configuration avoids unnecessary computational overhead. It is the most cost--effective combination of stencils in this work and is recommended for all subsequent simulations of STGV.

\begin{figure}[H]
    \centering
    \includegraphics[width=0.7\linewidth]{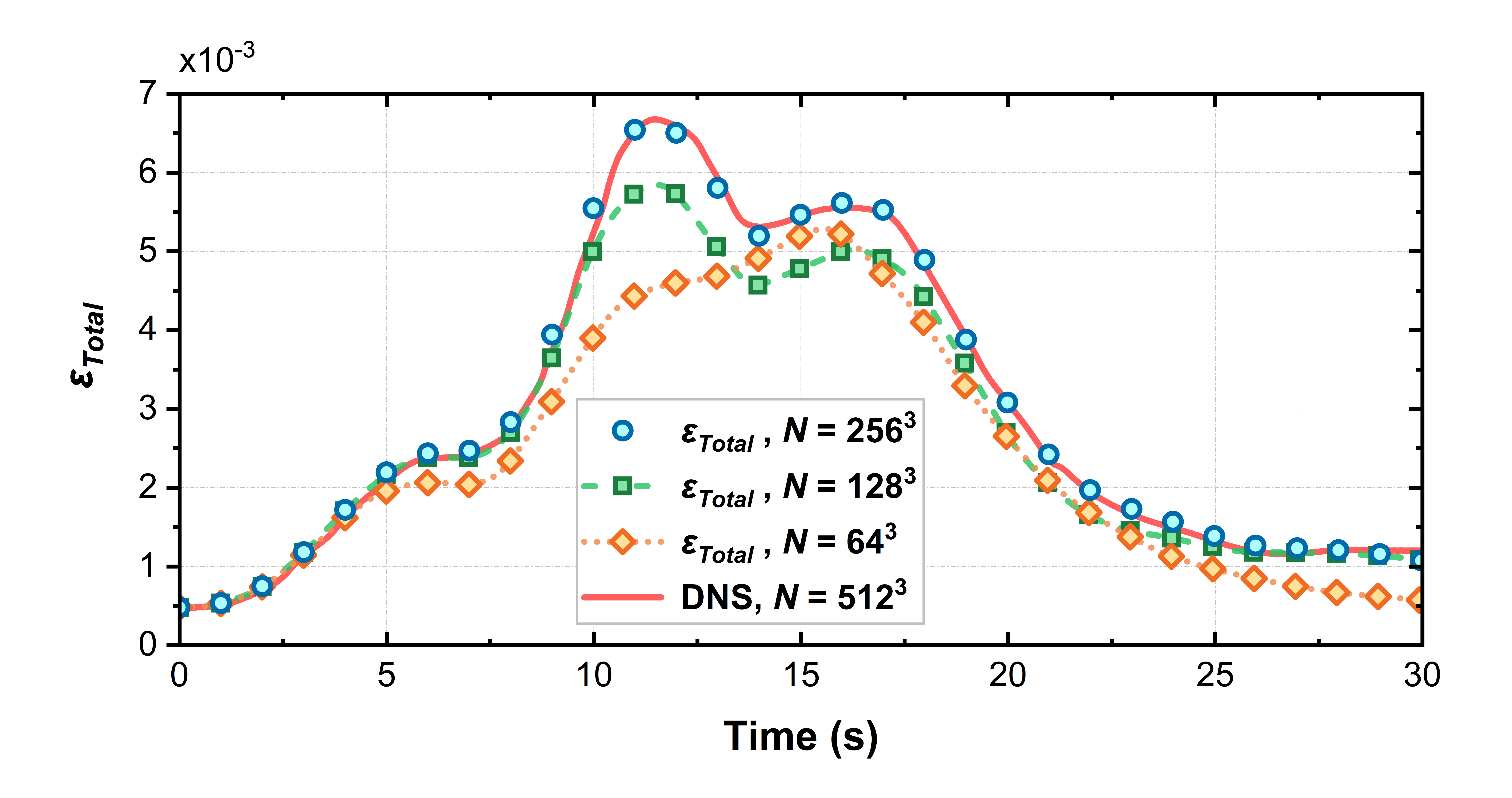}
    \caption{Effect of grid resolution on the total dissipation rate predicted by the D3Q27$\times$19 configuration. DNS from Jadhav et al.\cite{jadhav2021assessment}}
    \label{fig:3total}
\end{figure}

Having identified D3Q27$\times$19 configuration as the preferred combination, this section examines the impact of grid resolution in more detail. To determine the optimal grid resolution required to achieve the desired accuracy, the evolution curves of the total dissipation rate are compared in Fig.~\ref{fig:3total}. The kinetic and potential contributions to the dissipation rate are defined respectively as:
\begin{equation}
\varepsilon_k = \frac{\nu}{(2\pi)^3} \int_V (\nabla \times \mathbf{u})^2 \ dV,
\end{equation}

\begin{equation}
\varepsilon_p = \frac{\widetilde{D}}{(2\pi)^3} \int_V \left( \frac{G}{N} \frac{\nabla \rho'}{\rho_0} \right)^2 \, dV,
\end{equation}
 at three grid resolutions $N_g = 64^3, 128^3, 256^3$ and compared against the DNS data.
 
\begin{figure}[H]
    \centering
    \includegraphics[width=0.7\linewidth]{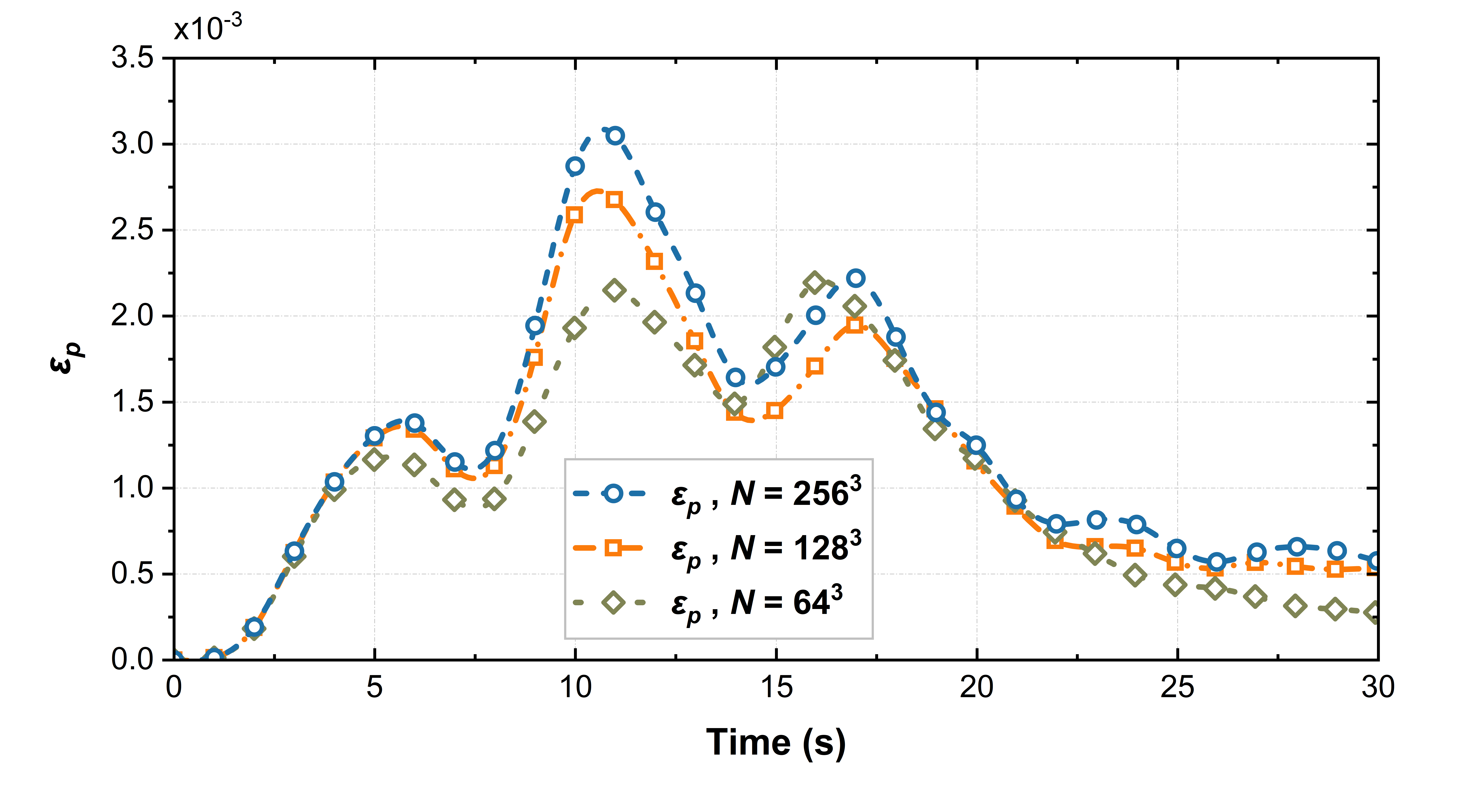}
    \caption{The time evolution of the potential energy dissipation rate $\varepsilon_p$ for three resolutions using the D3Q27$\times$19 stencil. DNS from Jadhav et al.}
    \label{fig:2}
\end{figure}

\begin{figure}[H]
    \centering
    \includegraphics[width=0.7\linewidth]{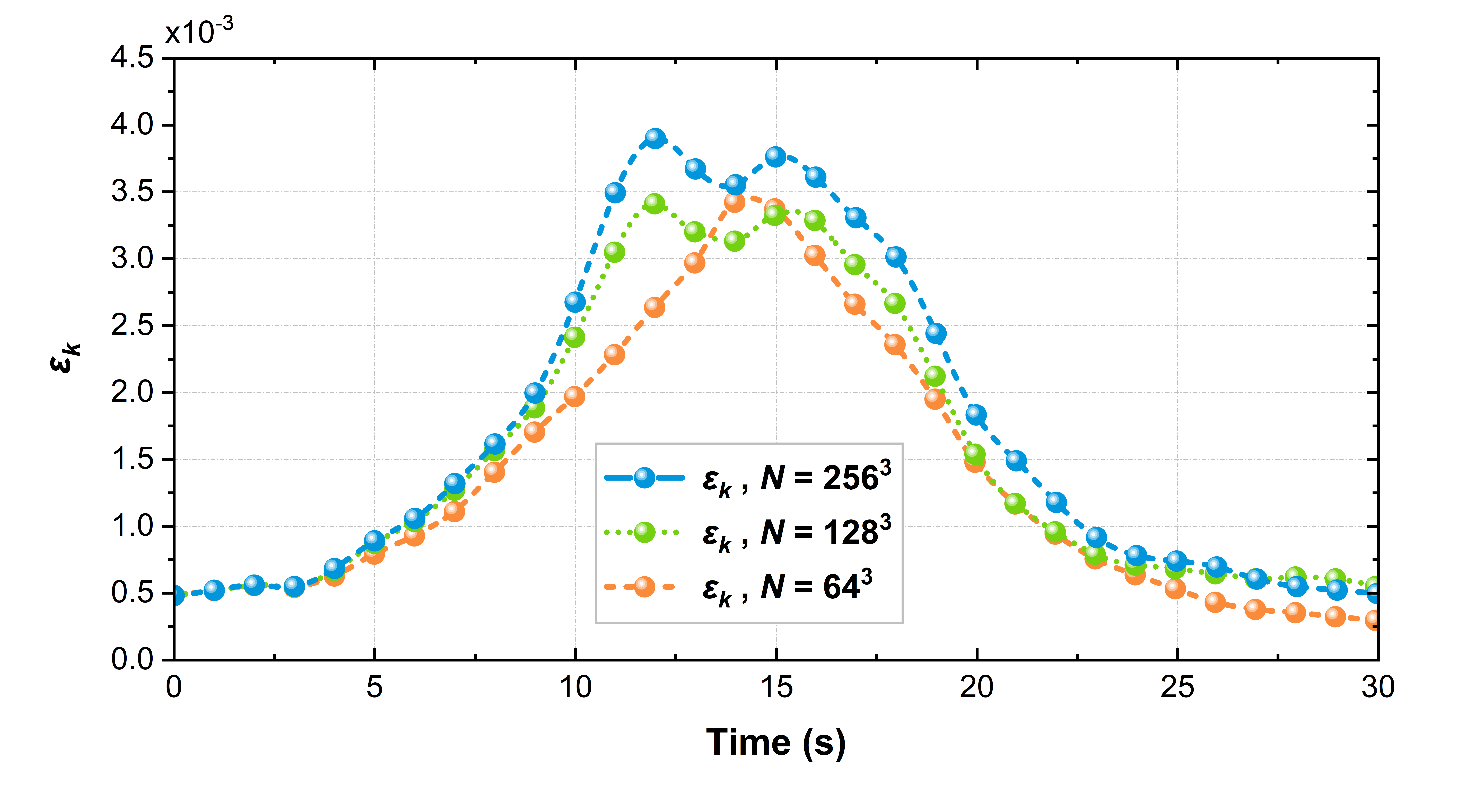}
    \caption{
    The time evolution of the kinetic energy dissipation rate $\varepsilon_k$ for three resolutions using the D3Q27$\times$19 stencil. DNS from Jadhav et al.\cite{jadhav2021assessment}}
    \label{fig:3}
\end{figure}

The sensitivity of the total dissipation rate to grid resolution varies significantly across different evolutionary stages of the flow. During the initial phase (Figs.~\ref{fig:2} and \ref{fig:3}), the flow field is dominated by large-scale ordered structures, yielding a low and steady dissipation rate. Because the dynamics are governed by these well-resolved macroscopic features, the LBM profiles for all grid resolutions perfectly collapse onto the DNS curve, demonstrating that early-stage dissipation is essentially grid-independent.

As the flow enters the double-dissipation-peak regime with the onset of fully developed turbulence, the sensitivity to grid resolution becomes highly pronounced. The coarse $64^3$ grid produces substantial errors in the amplitudes of the two dissipation peaks, capturing only 69.48\% of the corresponding DNS values. The overall dissipation curve is significantly underestimated and exhibits a smoother profile, indicating insufficient resolution of small-scale turbulent structures. As a result, energy transfer to the smallest resolved scales is prematurely damped by numerical viscosity. Increasing the resolution to $128^3$ leads to a marked improvement. The general trend and peak timing align closely with the DNS reference; however, the amplitude of the primary dissipation peak remains underpredicted by approximately 13.63\%. This residual discrepancy suggests that certain high-wavenumber components are still partially suppressed by numerical dissipation. The $256^3$ simulation provides the most accurate results, exhibiting excellent agreement with the DNS data over the entire time range. Both the peak magnitudes and the detailed temporal evolution in the double-peak regime are well reproduced. These results indicate that, under the present parameters, the total dissipation rate approaches grid convergence at a resolution of $256^3$.

During the late decay stage, the dissipation rate exhibits a monotonic decrease across all three grid resolutions. However, the profiles for the $64^3$ and $128^3$ cases remain consistently below the DNS reference, whereas the $256^3$ simulation maintains near-perfect agreement. This behaviour indicates that energy and dissipation errors introduced by insufficient grid resolution at earlier stages accumulate over time and subsequently affect the long-term flow evolution. Consequently, only the $256^3$ resolution can reliably resolve small-scale structures and accurately capture their contribution to the dissipation process throughout the entire evolution of the flow.


Based on the preceding analysis, a grid resolution of $N = 256^3$ is adopted as the minimum requirement to ensure quantitative precision when investigating the STGV with the D3Q27$\times$19 configuration. All subsequent parameter studies and analyzes of the underlying physical mechanisms are, therefore, conducted at this validated resolution.

\subsection{Buoyancy Reynolds number}

\begin{figure}[H]
    \centering
    \includegraphics[width=0.7\linewidth]{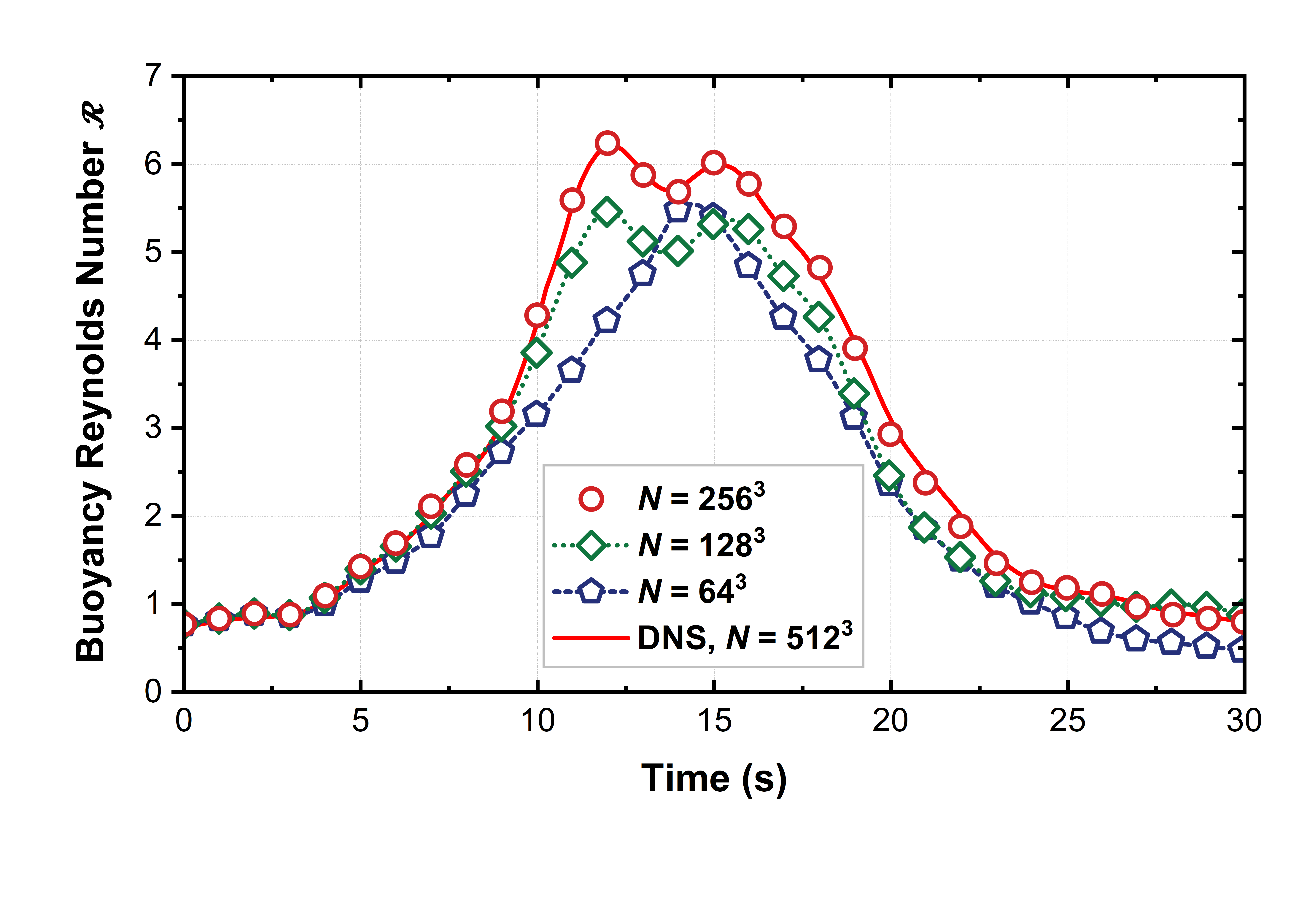}
    \caption{Effect of spatial resolution on the buoyancy Reynolds number $R$ predicted by the D3Q27$\times$19 configuration.}
    \label{fig:Buoyancy Reynolds number}
\end{figure}

To further assess the impact of grid resolution on small-scale turbulent structures, Fig.~\ref{fig:Buoyancy Reynolds number} presents the temporal evolution of the buoyancy Reynolds number $R$, defined as:
\begin{equation}
R= \frac{\varepsilon_k}{\nu N^2}.
\end{equation}
$R$ is a key dimensionless parameter in stratified turbulence, characterising the strength of nonlinear inertial effects relative to buoyancy forces. Its peak corresponds to the period of the most intense turbulent activity.

It is evident that the predictive accuracy of $R$ is strongly dependent on grid resolution. The coarse grid with $N = 64^3$ exhibits significant distortion, substantially underpredicting the peak value of $R$ by more than 12.22\% and showing a noticeable delay in the peak occurrence time. This behaviour indicates that excessive numerical dissipation suppresses the full development of turbulence at insufficient resolution. 

The intermediate grid with $N = 128^3$ captures the overall temporal trend of $R$, yet still fails to accurately resolve the strong turbulent fluctuations near the peak. The predicted peak values (5.45 and 5.32) remain lower than the DNS reference values (6.23 and 6.01), reflecting residual under-resolution of high-intensity turbulent motions.

In contrast, the $N = 256^3$ simulation shows excellent agreement with the DNS benchmark at $N = 512^3$. Both the peak magnitude and the subsequent decay phase are accurately reproduced. These results further confirm the conclusion drawn from the potential energy analysis: a grid resolution of at least $256^3$ is required to reliably capture the fine-scale turbulent structures and their interaction with stratification in the present flow regime.

\subsection{Froude number effects}

\begin{figure}[H]
    \centering
    \includegraphics[width=0.7\linewidth]{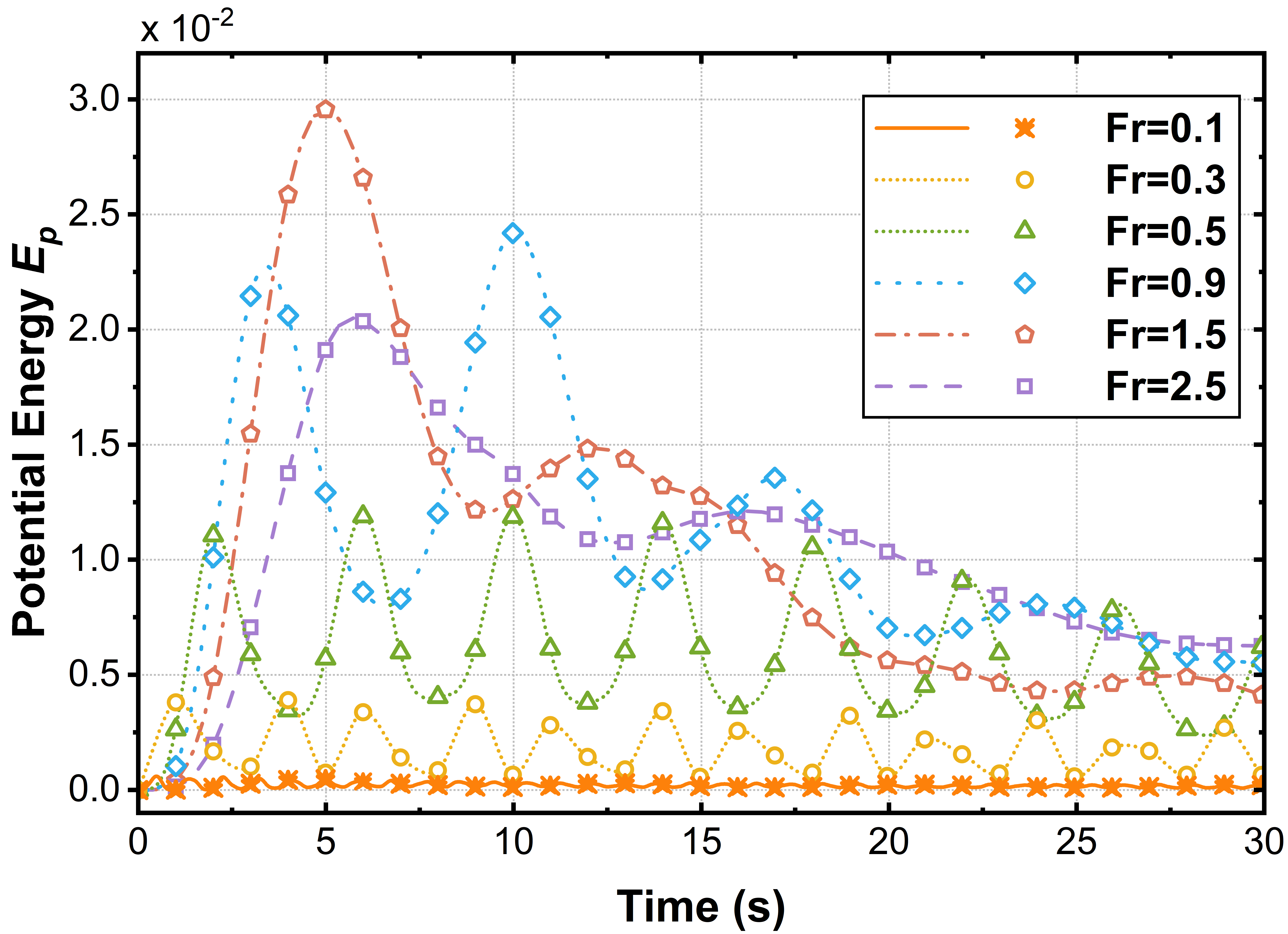}
    \caption{Potential energy over time for different Froude numbers at resolution $N_g = 256^3$ using the D3Q27$\times$19 stencils}
    \label{fig:FrEp}
\end{figure}

In stratified turbulence, the initial Froude number $Fr_0$ is a key dimensionless parameter that quantifies the ratio of inertial to buoyancy forces. When $Fr_0$ is small, buoyancy dominates the dynamics, and the flow exhibits strongly stratified behaviour characterised by internal--wave motions. As $Fr_0$ increases, stratification effects weaken and the flow progressively approaches the characteristics of unstratified turbulence. Consequently, $Fr_0$ plays a central role in the regulation of the exchange between kinetic and potential energy, the intensity of turbulent mixing, and the distribution of energy across spatial scales.
\begin{figure}[htbp]
    \centering
    \includegraphics[width=0.7\linewidth]{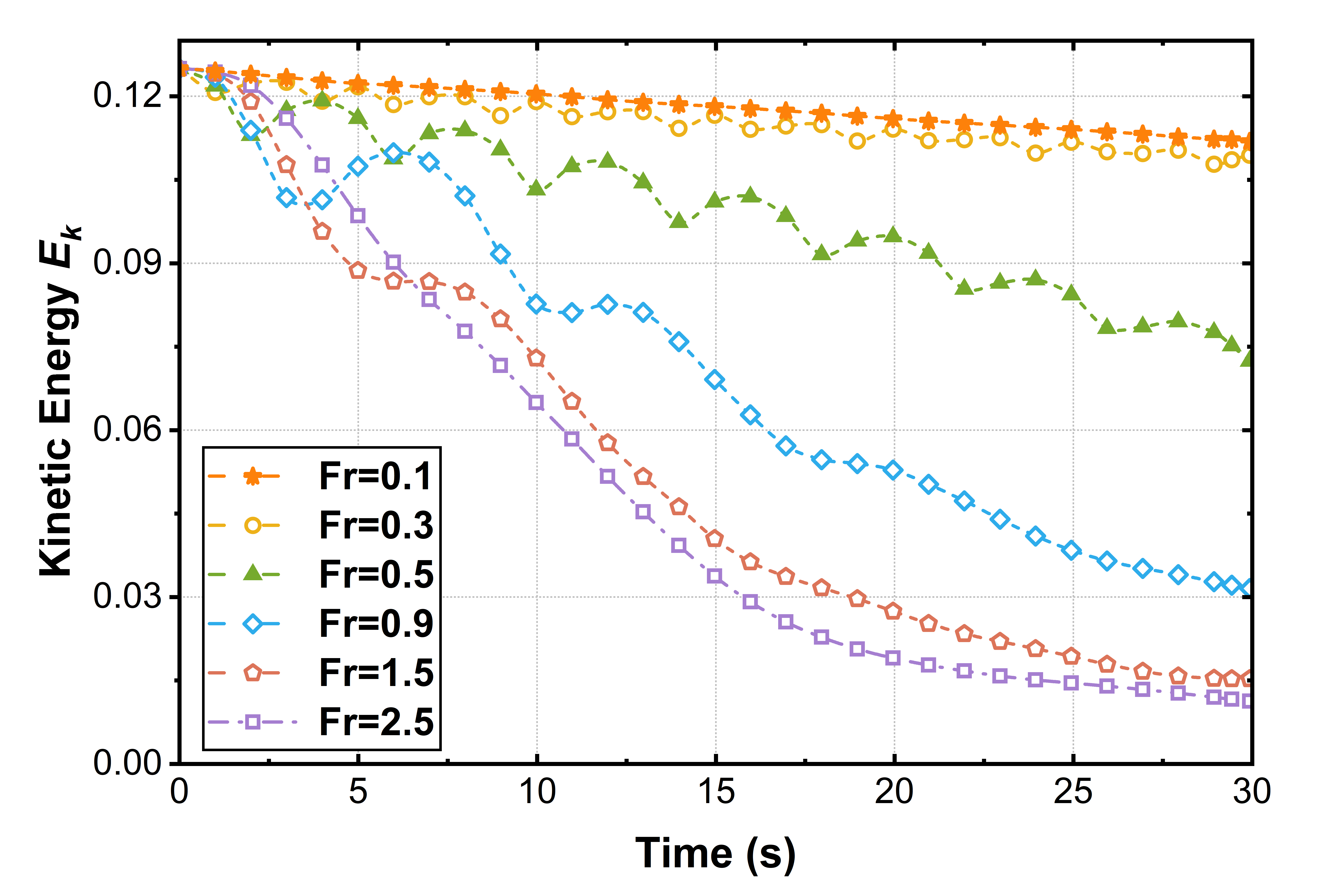}
    \caption{Kinetic energy over time for different Froude numbers at resolution $N_g = 256^3$ using the D3Q27$\times$19 stencils}
    \label{fig:FrEk}
\end{figure}
Figures~\ref{fig:FrEk} and~\ref{fig:FrEp} present the temporal evolution of the volume-averaged kinetic energy $E_k$ and potential energy $E_p$ for different initial Froude numbers $Fr_0$. As shown in Fig.~\ref{fig:FrEk}, stratification strength plays a key role in regulating the decay of $E_k$. Under strong stratification ($Fr_0 = 0.1$ and $0.3$), $E_k$ exhibits only weak decay and remains close to its initial value throughout the simulation. This indicates that buoyancy effects suppress nonlinear energy transfer and inhibit the cascade toward smaller scales. As $Fr_0$ increases to $0.5$ and $0.9$, the decay rate of $E_k$ progressively increases, reflecting the gradual weakening of buoyancy constraints. When stratification is further reduced ($Fr_0 = 1.5$ and $2.5$), $E_k$ decreases rapidly at early times, indicating that nonlinear turbulence and viscous dissipation dominate the flow dynamics. 

Figure~\ref{fig:FrEp} shows a corresponding oscillatory evolution of the potential energy $E_p$. In the strongly stratified regimes, $Fr_0 = 0.1$ and $0.3$, $E_p$ remains at a low level with only small-amplitude fluctuations, indicating that kinetic-to-potential energy transfer is strongly constrained. As $Fr_0$ increases to $0.5$ and $0.9$, pronounced peaks in $E_p$ emerge, reaching magnitudes on the order of $\mathcal{O}(10^{-2})$ and undergoing multiple oscillatory decay cycles. This behaviour reflects repeated energy exchange between kinetic and potential components mediated by internal-wave dynamics. The largest peaks occur at $Fr_0 = 0.9$, suggesting that energy conversion efficiency is maximised under moderate stratification. In the weakly stratified regimes ($Fr_0 = 1.5$ and $2.5$), $E_p$ is rapidly amplified during the early stage, followed by a gradual decay phase. This evolution corresponds closely to the rapid dissipation of kinetic energy observed in Fig.~\ref{fig:FrEk}.

\begin{figure}[H]
    \centering
    \includegraphics[width=0.7\linewidth]{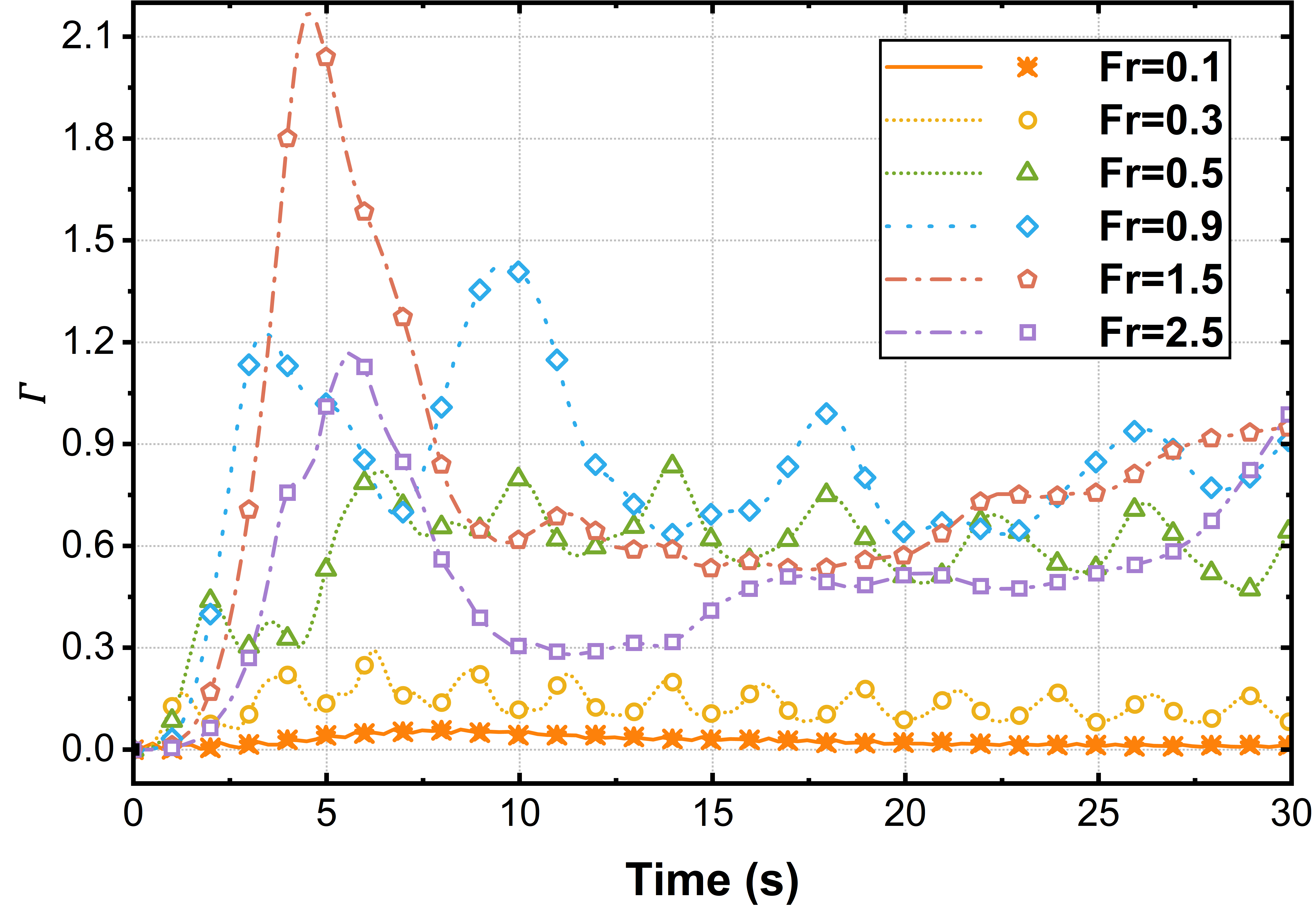}
    \caption{Flux coefficient $\Gamma$ for different Froude numbers.}
    \label{fig:Gamma_Time}
\end{figure}

Figure~\ref{fig:Gamma_Time} presents the temporal evolution of the flux coefficient $\Gamma = \varepsilon_p / \varepsilon_k$ for different initial Froude numbers $Fr_0$~\cite{onuki2021simulating,osborn1980estimates,mashayek2013time,vandine2021turbulent}. This coefficient provides a quantitative measure of mixing efficiency by characterising the relative dissipation of potential and kinetic energy.

Under strong stratification ($Fr_0 = 0.1$ and $0.3$), $\Gamma$ remains low throughout the simulation. For $Fr_0 = 0.1$, $\Gamma$ stays close to zero, whereas for $Fr_0 = 0.3$ it fluctuates within the range $0.06$--$0.29$. These values indicate that potential energy dissipation is substantially weaker than kinetic energy dissipation, resulting in limited mixing efficiency.

As $Fr_0$ increases to $0.5$ and $0.9$, $\Gamma$ rises rapidly during the turbulence development stage, reaching peak values of $0.81$ and $1.22$, respectively, before stabilising within the range $0.49$--$0.93$. This behaviour suggests that, under moderate stratification, the dissipation rates of kinetic and potential energy become comparable, thereby promoting enhanced energy conversion and mixing.

In the weakly stratified regimes ($Fr_0 = 1.5$ and $2.5$), the early-time peaks of $\Gamma$ increase further. For $Fr_0 = 1.5$, $\Gamma$ reaches approximately $2.16$ at $t \approx 4.48$, indicating that potential energy dissipation temporarily exceeds kinetic energy dissipation. For $Fr_0 = 2.5$, an initial peak of about $1.17$ is observed, followed by a decline to a minimum near $0.28$ and a gradual recovery.

Overall, the peak value of $\Gamma$ increases significantly as $Fr_0$ transitions from strong to moderate stratification and then stabilises or decreases in the weakly stratified regime. These results demonstrate that stratification strength governs not only the absolute energy levels but also their relative dissipation pathways. The dependence of mixing efficiency on $Fr_0$ is therefore distinctly non-monotonic, with maximum efficiency occurring under moderate stratification and strong suppression in the strongly stratified limit.

Next, the bulk flux coefficient $\Gamma_b$ is defined as the ratio of the time-integrated potential energy dissipation to the time-integrated kinetic energy dissipation:
\begin{equation}
\Gamma_b = \frac{\int_0^{30} \varepsilon_p \, dt}{\int_0^{30} \varepsilon_k \, dt}.
\end{equation}
This coefficient provides a global measure of mixing efficiency over the entire flow evolution.

\begin{figure}[H]
    \centering
    \includegraphics[width=0.7\linewidth]{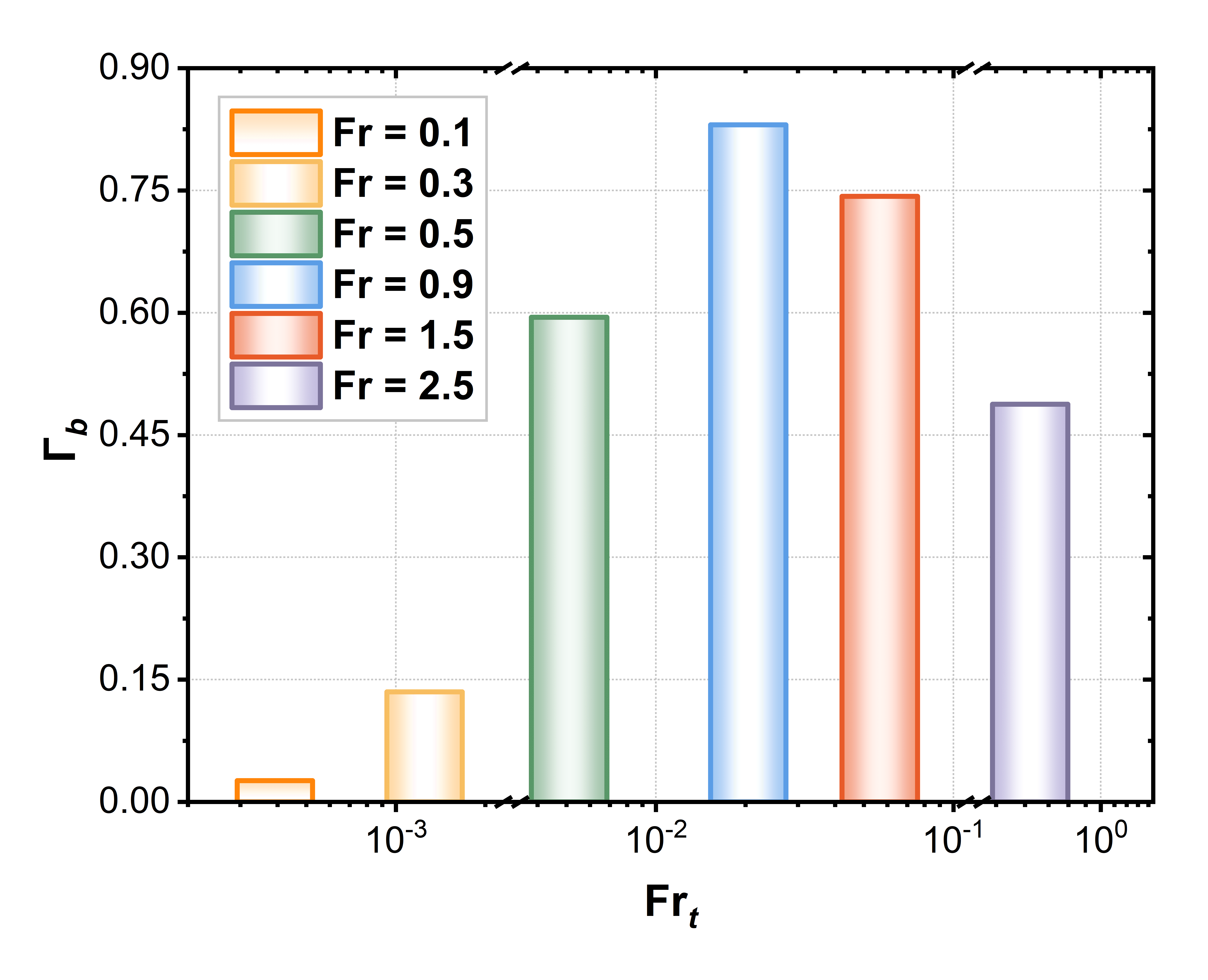}
    \caption{ Relation of flux coefficient of the bulk $\Gamma_b$ with time-integrated turbulent Froude number $Fr_0$ for different initial Froude numbers.}
    \label{fig:frrb}
\end{figure}

Figure~\ref{fig:frrb} presents the relationship between $\Gamma_b$ and the turbulent Froude number $Fr_t$ for different initial Froude numbers $Fr_0$. The turbulent Froude number, representing the ratio of inertial to buoyancy forces in the bulk, is defined as
\begin{equation}
\mathrm{Fr}_t = \frac{\int_0^{30} \varepsilon_k \, dt}{N \int_0^{30} E_k \, dt}.
\end{equation}
As $Fr_t$ increases, $\Gamma_b$ exhibits a pronounced non-monotonic dependence. Under very strong stratification, $\Gamma_b$ remains low, indicating that potential energy dissipation is substantially smaller than kinetic energy dissipation and that the mean mixing efficiency is limited. As $Fr_0$ increases into the moderately stratified regime, $\Gamma_b$ rises rapidly and attains a maximum in the range $0.59$--$0.83$. This behaviour reflects a more balanced time-integrated dissipation of kinetic and potential energy, corresponding to enhanced overall energy conversion and mixing.

When stratification is further weakened (e.g., $Fr_0 = 2.5$), $\Gamma_b$ decreases to approximately $0.48$. This indicates that, under near-unstratified conditions, a larger fraction of energy is dissipated directly through the kinetic pathway, while the proportion converted into potential energy is reduced. To further elucidate the scale-dependent mechanisms underlying this non-monotonic behaviour, the kinetic energy spectra are analysed below.

\begin{figure}[htbp]
    \centering
    \includegraphics[width=0.7\linewidth]{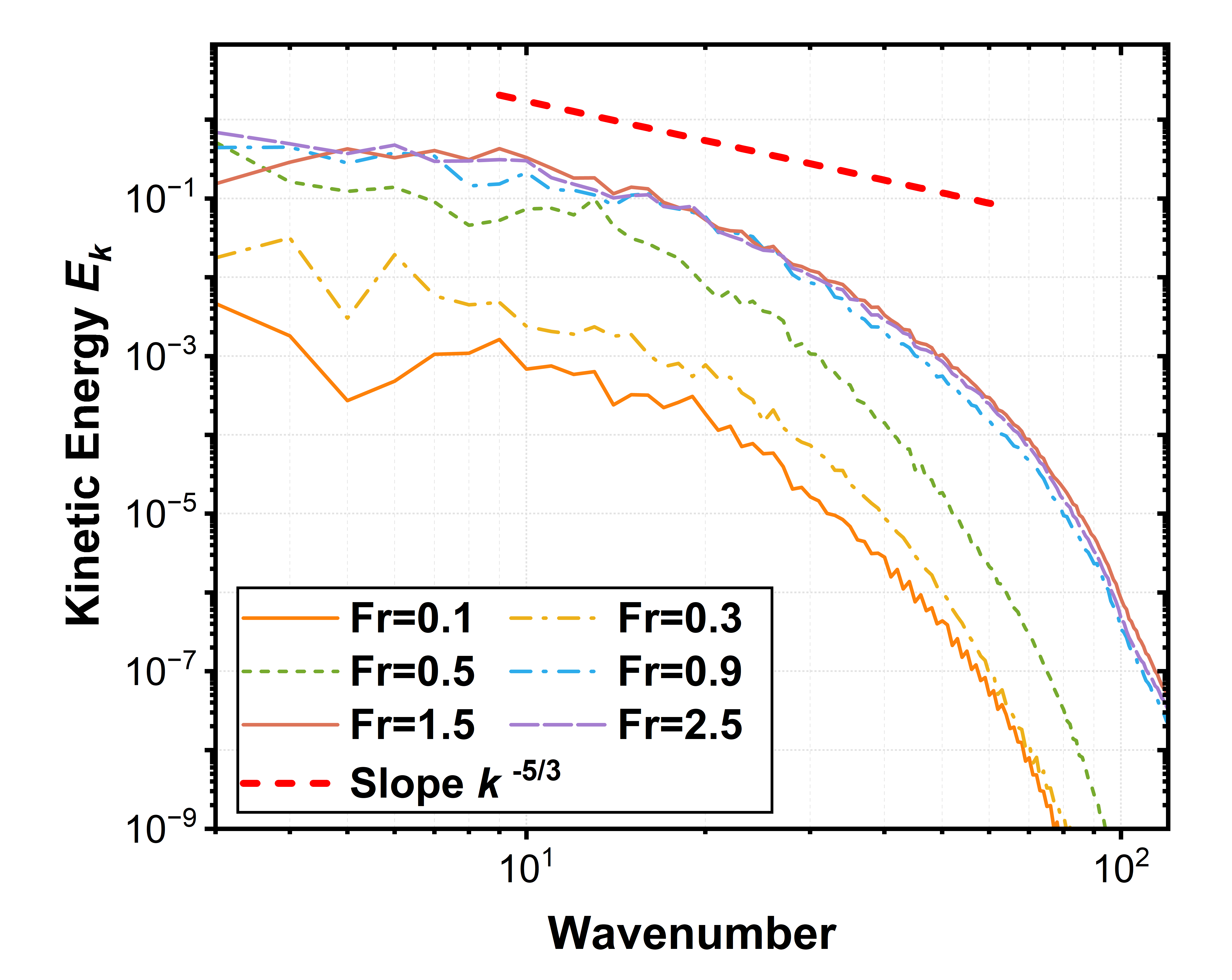}
    \caption{ Kinetic energy over wavenumber at t= 15 for the same Froude numbers. The dashed line shows the slope $k^{-5/3}$. }
    \label{fig:ekwavenumber}
\end{figure}

Figure~\ref{fig:ekwavenumber} presents the kinetic energy spectra at peak turbulent activity for different initial Froude numbers $Fr_0$. The horizontal axis denotes the wavenumber $k$, while the vertical axis represents the kinetic energy spectral density. A reference slope of $k^{-5/3}$ is included to indicate the expected inertial-subrange scaling.

Stratification strength markedly influences the spectral distribution of energy across scales. In the strongly stratified cases ($Fr_0 = 0.1$ and $0.3$), the spectra remain significantly lower than those of other cases over the entire wavenumber range, with particularly rapid decay at intermediate and high wavenumbers. This behaviour demonstrates that strong stratification suppresses the forward cascade of kinetic energy toward small scales, leading to flow dynamics dominated by large-scale internal waves and layered structures.

As $Fr_0$ increases to $0.5$ and above, the spectra corresponding to moderately and weakly stratified regimes progressively collapse in the intermediate wavenumber range. Their amplitudes become comparable and exhibit approximate inertial-subrange scaling close to the reference $k^{-5/3}$ slope, indicating more efficient energy transfer to small scales and enhanced small-scale turbulence. In particular, the spectra for $Fr_0 = 0.9$, $1.5$, and $2.5$ nearly overlap at intermediate and high wavenumbers, suggesting that the small-scale dynamics approach those of nearly isotropic turbulence under moderate and weak stratification.

Overall, the spectral characteristics provide scale-resolved support for the trends identified in the energy evolution and mixing analyses. Strong stratification (small $Fr_0$) suppresses the energy cascade, resulting in weakened small-scale turbulence and reduced mixing efficiency. In contrast, moderate to weak stratification sustains a forward cascade toward small scales and promotes approximate $k^{-5/3}$ scaling, consistent with the regime exhibiting the highest mixing activity.

\begin{figure}[H]
    \centering
    \includegraphics[width=0.7\linewidth]{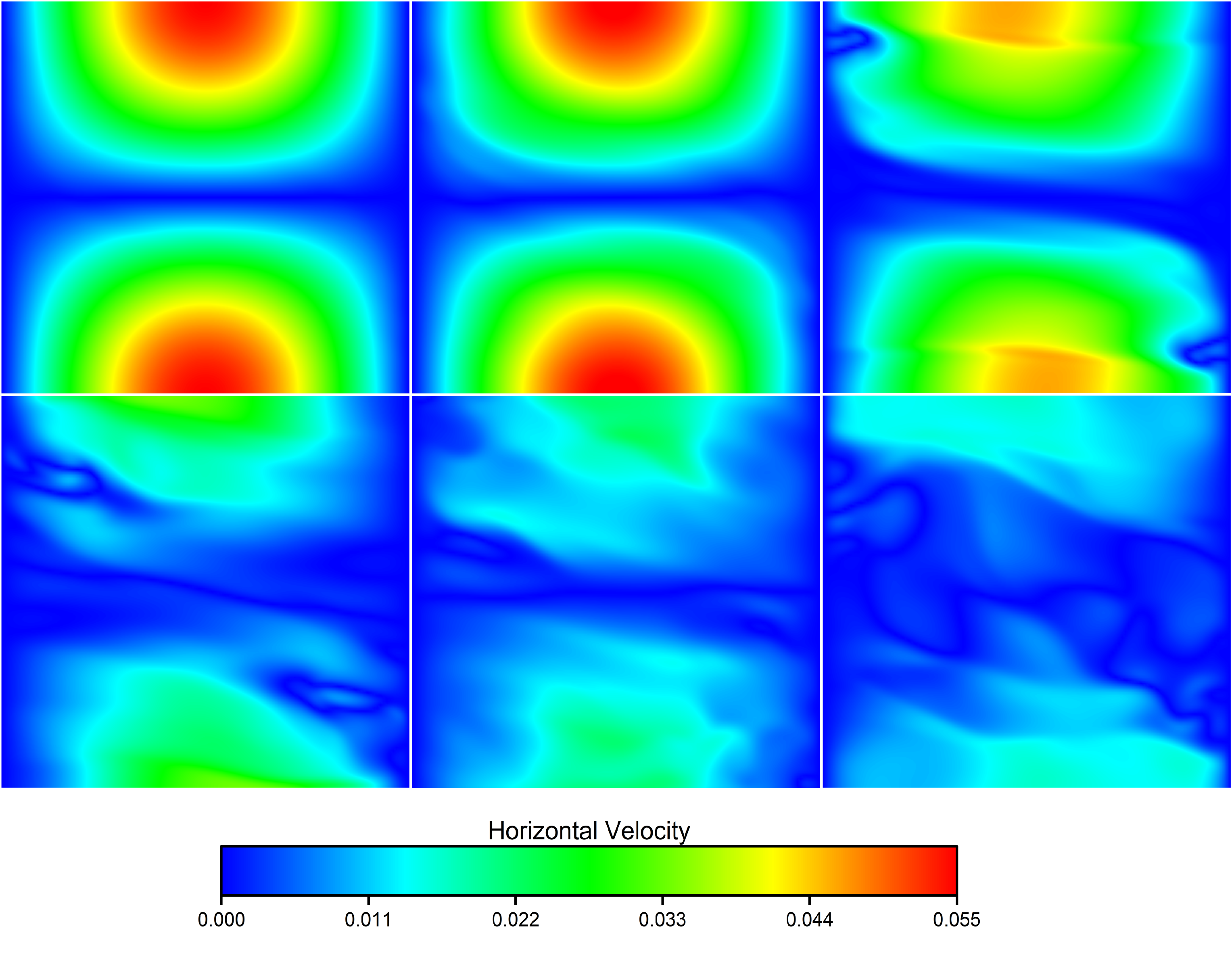}
    \caption{Horizontal velocity of the y-z plane at x = 0 and t = 15 for different Froude numbers.}
    \label{fig:hjm}
\end{figure}

Figure~\ref{fig:hjm} further illustrates the influence of $Fr_0$ on the horizontal velocity magnitude $\sqrt{u_x^2 + u_y^2}$. Under strong stratification (e.g., $Fr_0 = 0.3$), coherent large-scale structures are preserved, indicating suppressed small-scale breakdown and limited turbulent mixing.

In summary, the initial Froude number $Fr_0$ fundamentally controls the energy transfer pathways and mixing dynamics of the stratified Taylor–Green vortex. Strong stratification suppresses the forward cascade of kinetic energy, limits kinetic-to-potential energy conversion, and weakens small-scale turbulence, resulting in low mixing efficiency. As stratification weakens, energy exchange between kinetic and potential components becomes more effective, and a sustained cascade toward small scales develops. Both the time-dependent and time-integrated flux coefficients demonstrate a distinctly non-monotonic dependence on $Fr_0$, with maximum mixing efficiency occurring under moderate stratification. The corresponding energy spectra and flow structures further confirm that intermediate buoyancy effects provide the most favourable conditions for sustained turbulent mixing.

\subsection{Comparison of forcing schemes}

To assess the numerical accuracy and suitability of different lattice Boltzmann forcing formulations for stratified flow simulations, this subsection systematically compares four classical schemes: the Shan-Chen (SC) model \cite{shan1993lattice}, the Exact Difference Method (EDM) \cite{kupershtokh2009equations}, the formulation of Guo et al. (Guo) \cite{guo2002discrete}, and the HCZ scheme \cite{he1998discrete}. The normalized kinetic energy evolution, $E_k/E_{k0}$, is employed as the primary diagnostic quantity to evaluate their robustness across varying stratification intensities.

\begin{figure}[H]
    \centering
    \includegraphics[width=0.7\linewidth]{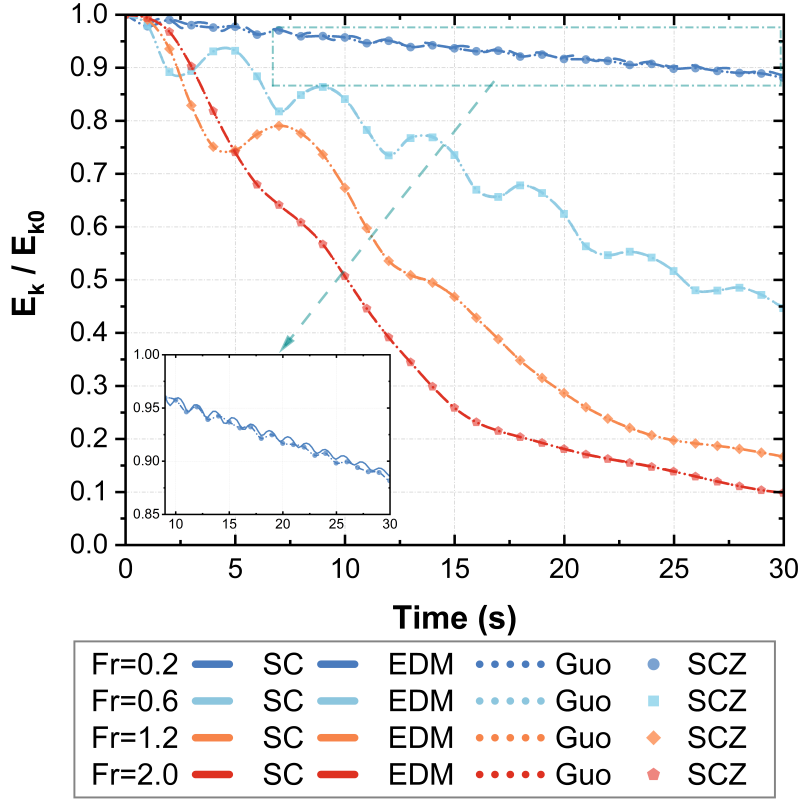}
    \caption{ Time evolution of the normalized kinetic energy $E_k/E_{k_0}$ for different Froude numbers $Fr_0$ and forcing schemes. }
    \label{fig:Force}
\end{figure}

As illustrated in Fig.~\ref{fig:Force}, the sensitivity of the kinetic energy evolution to the forcing formulation depends strongly on the initial Froude number $Fr_0$. In the high-$Fr_0$ regime, all four schemes produce nearly indistinguishable results, indicating that force discretization errors remain negligible when inertial effects dominate. However, as $Fr_0$ decreases and buoyancy effects become increasingly influential, pronounced discrepancies emerge among the formulations. This divergence indicates that the mathematical discretization of the external force plays a decisive role in accurately capturing the energy evolution under strong stratification.

\begin{figure}[H]
    \centering
    \includegraphics[width=0.7\linewidth]{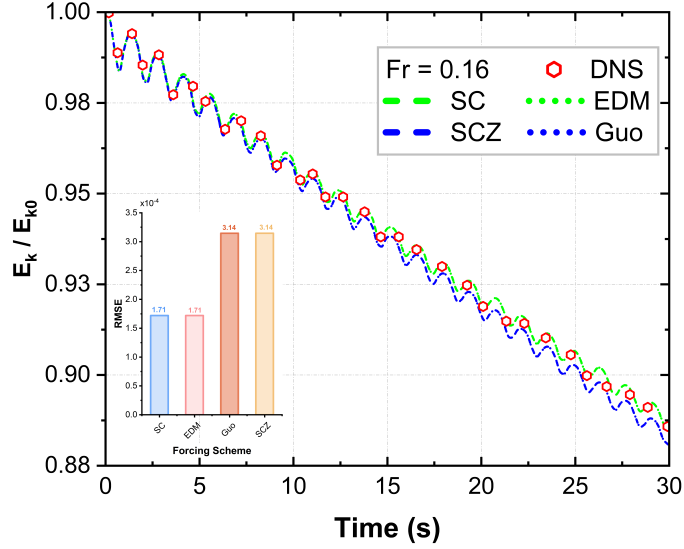}
    \caption{Time evolution of the normalized kinetic energy $E_k/E_{k_0}$
at $Fr=0.16$ with different forcing schemes, compared with the DNS reference.
The inset shows the corresponding RMSE.}\label{fig:0.16}
\end{figure}

To quantitatively identify the most suitable formulation in strongly stratified conditions, we focus on the extreme case $Fr_0 = 0.16$ and benchmark the LBM results against DNS data. As shown in Fig.~\ref{fig:0.16}, all schemes successfully reproduce the fundamental oscillatory behaviour and overall decay of $E_k$. Nevertheless, the results bifurcate into two distinct clusters. The SC and EDM formulations are virtually congruent and exhibit excellent agreement with the DNS reference. In contrast, the Guo and HCZ formulations largely overlap but exhibit a noticeable deviation from the DNS benchmark.

For a more rigorous comparison, the Root Mean Square Error (RMSE) relative to DNS is computed:
\begin{equation}
\text{RMSE} = \sqrt{\frac{1}{N}\sum_{i=1}^{N}\left(E_k^\text{Sim} - E_k^\text{DNS}\right)^2}
\end{equation}
where $E_k^{\mathrm{Sim}}$ and $E_k^{\mathrm{DNS}}$ denote the simulated and DNS kinetic energy, respectively, and $N$ represents the number of sampling points. As indicated in the bar chart (inset of Fig.~\ref{fig:0.16}), the SC and EDM formulations yield the lowest RMSE values, whereas the Guo and HCZ schemes exhibit comparatively larger errors. Within the present numerical framework, the velocity-shift formulations (SC and EDM) lead to an approximately 45.33\% reduction in the RMSE relative to the discrete source term approaches.
The observed clustering behaviour is rooted in the theoretical structure of the forcing formulations. The SC and EDM schemes belong to the velocity-shift family, with EDM representing the exact Taylor expansion of the SC formulation. Under the incompressible approximation, their macroscopic momentum equations are essentially equivalent, leading to nearly identical numerical trajectories. By contrast, the Guo and HCZ schemes are categorized as discrete source term formulations, which introduce explicit force source terms and redefined macroscopic velocities to achieve second-order temporal accuracy. Their similar mathematical frameworks account for the high degree of overlap in their results.

Notably, while the Guo and HCZ schemes are theoretically designed to minimize discrete lattice effects, the SC and EDM formulations demonstrate superior fidelity in the strongly stratified regime. This is attributed to the fact that stratified shear flows are exceptionally sensitive to numerical dissipation. In Guo-type schemes, the explicit half-step force correction may introduce a subtle pseudo-viscosity under strong external force gradients, accelerating $E_k$ decay beyond the physical benchmark. Conversely, by coupling the force contribution directly into the equilibrium relaxation process, the SC and EDM schemes benefit from a favourable error-cancellation effect between their inherent truncation errors during periodic vortex evolution, thereby preserving the system's physical dissipation characteristics with higher precision.

In summary, while the choice of forcing formulation has negligible influence in inertia-dominated regimes, it becomes a critical factor in low-$Fr_0$ stratified flows. The SC and EDM velocity-shift formulations exhibit superior numerical fidelity under strong buoyancy effects and are therefore adopted for the remainder of this study.

\subsection{Analysis of symmetry--breaking velocity errors}
While macroscopic quantities such as kinetic energy provide a useful assessment of global flow behaviour, they may not fully reveal subtle numerical artefacts arising from lattice anisotropy or force discretisation. In strongly stratified flows, small symmetry violations can accumulate over time and manifest as spurious parasitic currents that do not significantly affect global energy decay but compromise the microscopic force balance. Therefore, a more stringent diagnostic based on symmetry preservation is required to evaluate the intrinsic numerical fidelity of different stencil configurations.

\begin{figure}[H]
    \centering
    \includegraphics[width=0.7\linewidth]{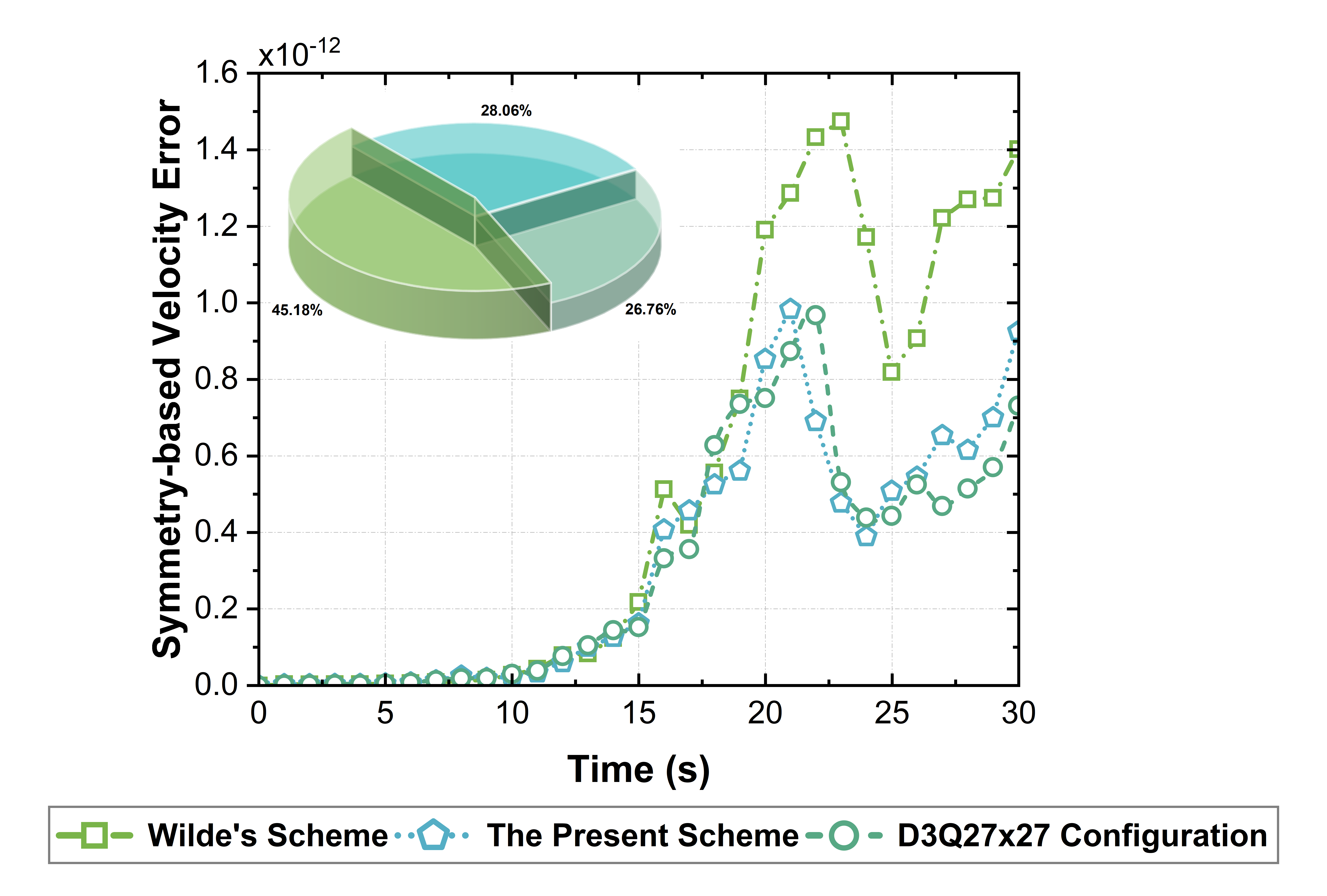}
    \caption{ Time evolution of the symmetry-breaking velocity error for different stencil combinations. The inset pie chart shows the cumulative error contribution of each case. }
    \label{fig:Error_Total}
\end{figure}

As shown in Fig.~\ref{fig:Error_Total}, all schemes initially maintain symmetry with negligible error. However, as nonlinear interactions intensify ($t > 15$ s), measurable deviations emerge. Wilde’s scheme exhibits the largest symmetry-breaking error throughout the later stages, with peak values exceeding those of the other configurations. Although Wilde’s formulation effectively reproduces the overall kinetic energy decay, its substantial accumulated symmetry error (accounting for 45.18\% of the total error, as shown in the inset) indicates reduced fidelity in preserving microscopic force balance under strong stratification.

Notably, the proposed configuration (D3Q27$\times$19) demonstrates a marked improvement in symmetry preservation. Its error profile aligns closely with that of the full D3Q27$\times$27 configuration, which serves as the highest-order lattice configuration considered in this study. The cumulative error distribution confirms this precision saturation: the present scheme contributes only 26.76\% of the total error, nearly matching the 28.06\% observed for the more computationally demanding D3Q27$\times$27 model. This suggests that the D3Q19 scalar stencil provides sufficient lattice isotropy to maintain high compatibility with the D3Q27 momentum solver, thereby suppressing the growth of spurious parasitic flows.

In summary, while Wilde’s configuration remains capable of capturing global trends, the present scheme achieves improved long-term numerical fidelity by substantially reducing symmetry-breaking errors. By reaching precision saturation with a 19-speed scalar stencil, the proposed configuration offers an optimal balance between numerical accuracy and computational efficiency for stratified turbulence simulations.

\section{Conclusion}
\section*{Conclusions}

This work systematically investigates the performance of the lattice Boltzmann method (LBM) in simulating three-dimensional stably stratified Taylor–Green vortex (STGV) flows. Particular emphasis is placed on the influence of stencil combinations, forcing formulations, grid resolution, and Froude number on numerical accuracy, stability, and long-term fidelity. The results demonstrate that simulation quality critically depends on the consistent matching of velocity stencil isotropy, spatial resolution, and force discretisation.

Among the coupled stencil configurations examined, the D3Q27$\times$19 combination achieves the most favourable balance between accuracy and computational efficiency. It accurately reproduces the temporal evolution of kinetic energy, potential energy, and total dissipation rates, while maintaining low error accumulation during long-time integration. In contrast, lower-order momentum stencils such as D3Q7 introduce substantial numerical dissipation and spurious velocities. Even when paired with higher-order scalar stencils, such reduced-order momentum discretisations fail to preserve hydrostatic balance and accurately capture energy transfer mechanisms. Although the fully high-order D3Q27$\times$27 configuration offers maximal lattice symmetry, it does not provide substantial improvement over D3Q27$\times$19 in the present benchmark. This indicates that simply increasing stencil order does not necessarily yield proportional gains in overall numerical performance.

Grid and parameter sensitivity analyses further reveal that potential energy evolution and small-scale turbulent structures are considerably more sensitive to spatial resolution than kinetic energy. A grid resolution of $256^3$ is necessary to accurately resolve internal gravity wave dynamics and the characteristic double-peak structure of the dissipation rate. The Froude number analysis demonstrates that strong stratification suppresses vertical motion and inhibits the forward cascade of energy toward smaller scales, whereas moderate stratification yields the highest mixing efficiency and most effective energy exchange between kinetic and potential components.

Comparative analysis of external forcing formulations shows that, under strongly stratified conditions, the SC and EDM velocity-shift schemes consistently outperform the Guo and SCZ discrete source term approaches in terms of agreement with DNS benchmarks. The symmetry-breaking velocity error analysis further confirms that momentum-field isotropy is fundamental to numerical stability. In particular, the D3Q27$\times$19 configuration reaches a precision saturation point, maintaining symmetry preservation and suppressing spurious parasitic flows at a level comparable to the fully high-order D3Q27$\times$27 scheme, but at substantially reduced computational cost.

Overall, this study confirms the reliability and accuracy of LBM for simulations of stably stratified turbulence when appropriate discretisation strategies are employed. The isotropy of the velocity stencil, adequate grid resolution, and consistent force discretisation are identified as the primary determinants of simulation quality. Based on the comprehensive evaluation presented herein, the recommended configuration for the STGV benchmark consists of the D3Q27$\times$19 stencil combination, a $256^3$ grid resolution, and SC/EDM forcing schemes. These findings provide practical guidance for high-fidelity LBM simulations of stratified turbulent flows.

Future work may extend this numerical framework to more complex stratified turbulence scenarios, including higher Reynolds number regimes, multi-physics couplings, and applications in geophysical fluid dynamics.

\section{Acknowledgments}

This work was primarily supported by the National Natural Science Foundation of China (Grant No. 11872382) and the Thousand Young Talents Program of China.

The author sincerely acknowledges the guidance and generous support of Prof. Ewe-Wei Saw and Prof. Dominik Wilde, whose insightful discussions and continuous encouragement were invaluable to the completion of this work.

The author also thanks Fanxi Gong, Yating Chen, and Linli Fu for their helpful discussions and assistance throughout this work. Special thanks are extended to Ms. Li Zhang for her dedicated administrative and logistical support.
\bibliographystyle{unsrt}
\bibliography{Reference}

\end{document}